\documentclass[twocolumn,showpacs,preprintnumbers,amsmath,amssymb,superscriptaddress]{revtex4-1}

\usepackage{graphicx}
\usepackage{dcolumn}
\usepackage{bm}
\usepackage{amsthm,amsmath}

\usepackage[x11names]{xcolor}
\usepackage{longtable}
\usepackage{tikz}
\usepackage[T1]{fontenc}
\usepackage{lmodern}
\usepackage{tabularx}
\usepackage{longtable}
\usepackage{array}
\usepackage{multirow}

\tikzstyle{format}=[thin, rounded corners=3pt, line width =0.5, fill=gray!20!white, font=\scriptsize]
\tikzstyle{pseudo}=[font=\small]
\usetikzlibrary{patterns, mindmap, trees, positioning, shadings, fadings, calc, backgrounds}

\colorlet{mindmap1}{SpringGreen4}
\colorlet{mindmap2}{SpringGreen4!85}
\colorlet{mindmap3}{SpringGreen4!70}
\definecolor{elements}{RGB}{53,0,53}
\definecolor{binaries}{RGB}{0,153,0}

\begin{document}


\title[MSE]{Massive-parallel Implementation of the Resolution-of-Identity Coupled-cluster Approaches in the Numeric Atom-centered Orbital Framework for Molecular Systems}

\author{Tonghao Shen}
\affiliation{Fritz-Haber-Institut der Max-Planck-Gesellschaft, Faradayweg 4-6, 14195 Berlin, Germany}%
\affiliation{Department of Chemistry, Fudan University, Shanghai 200433, China}%

\author{Igor Ying Zhang}
\email{igor_zhangying@fudan.edu.cn}
\affiliation{Fritz-Haber-Institut der Max-Planck-Gesellschaft, Faradayweg 4-6, 14195 Berlin, Germany}%
\affiliation{Department of Chemistry, Fudan University, Shanghai 200433, China}%
\affiliation{MOE Key Laboratory of Computational Physical Science, Fudan University, Shanghai 200433, China}%

\author{Matthias Scheffler}
\affiliation{Fritz-Haber-Institut der Max-Planck-Gesellschaft, Faradayweg 4-6, 14195 Berlin, Germany}%

\date{\today}

\begin{abstract}
We present a massive-parallel implementation of the resolution-of-identity (RI) coupled-cluster approach 
that includes single, double and perturbatively triple excitations, namely RI-CCSD(T), in the FHI-aims package 
for molecular systems. A domain-based distributed-memory algorithm in the MPI/OpenMP hybrid framework 
has been designed to effectively utilize the memory bandwidth and significantly minimize the interconnect 
communication, particularly for the tensor contraction in the evaluation of the 
particle-particle ladder term. Our implementation features a rigorous avoidance of the on-the-fly disk 
storage and an excellent strong scaling up to 10,000 and more cores. Taking a set of molecules with different sizes, 
we demonstrate that the parallel performance of our CCSD(T) code is competitive with the CC implementations in 
state-of-the-art high-performance computing (HPC) computational chemistry packages. 
We also demonstrate that the numerical error due to the use of RI approximation in our RI-CCSD(T) is negligibly small.
Together with the correlation-consistent numeric atom-centered orbital (NAO) basis sets, 
NAO-VCC-$n$Z, the method is applied to produce accurate theoretical reference data for 22 bio-oriented weak interactions (S22), 
11 conformational energies of gaseous cysteine conformers (CYCONF), and 32 isomerization energies (ISO32).

\end{abstract}

\pacs{Valid PACS appear here}
\keywords{electronic-structure theory, density-functional theory,
second order perturbation theory, random-phase approximation}
\maketitle

\section{Introduction}
\label{introduction}

Coupled-cluster (CC) theory is a well-established wave function-based electronic-structure approach
that originates in nuclear physics~\cite{coester:1958,coester:1960A}, but flourishes in the 
quantum-chemistry community~\cite{cizek:1966A,paldus:1972A,szabo:1996A,bartlett:2007A}. Coupled-cluster 
theory holds many theoretical advantages, for example size extensivity~\cite{bartlett:2007A,hirata:2011A} 
and orbital invariance~\cite{bartlett:2007A,szabo:1996A}, both of which are crucial for a correct 
description of large and even extended systems. Compared to the widely 
used density-functional approximations, the CC hierarchy provides a systematic way to 
approach the exact description of the electron-correlation effects at least for systems with not too small HOMO-LUMO gaps. The coupled-cluster ansatz with 
single and double excitations (CCSD)~\cite{purvis:1982A} with its perturbative consideration of 
triple excitations, known as CCSD(T)~\cite{raghavachari:1989A}, has long been recognized ``gold standard''
in quantum chemistry. However, the price to pay for these potential benefits is a considerably increased 
numerical complexity, which manifests itself in an $O(N^6)$ computational scaling and an $O(N^4)$ memory 
requirement with system size ``$N$'' for the CCSD method based on the optimal formulation proposed 
by Scuseria, Janssen, and Schaefer~\cite{scuseria:1988A}. CCSD(T) shares the same memory requirement but with 
one order of magnitude higher computational scaling $O(N^7)$. This ``curse of dimensionality''\cite{bellman:1961A}
together with the well-known slow basis-set convergence problem~\cite{dunning:1989A,tew:2007A} significantly 
hamper the precise numerical computation of the CC methods to large systems.

Serveral reduced-scaling approximations of CC methods having been proposed for molecular 
systems~\cite{saebo:1985A,scuseria:1999A,schutz:2001A,riplinger:2016A}, but a high-performance 
massive-parallel implementation of 
conventional CCSD and CCSD(T) methods -- the goal of this paper -- is still highly valuable. It also
severs the ultimate benchmark for moderate-size systems in the complete basis set (CBS) limit for the 
reduced-scaling formalisms~\cite{peng:2016A} and it paves the way to study the applicability of the CC 
methods and their reduced-scaling variants in solids, which is emerging quickly as an active field in 
computational materials science~\cite{booth:2013A,michaelides:2015A,Gruneis:2017A}.

The first massive-parallel implementation of conventional closed-shell CCSD energy was reported 
in 1992 by Rendell \emph{et. al.}~\cite{rendell:1992A} adopting Scuseria's formulation with optimal 
$O(N^6)$ computational scaling~\cite{scuseria:1988A}. Rendell's algorithm is the base of most 
state-of-the-art CC codes aiming at massive-parallel calculations. It has been well 
recognized that, the major challenge towards an efficient massive-parallel CCSD(T) implementation 
based on Rendell's algorithm is not about the evaluation of the most expensive perturbative (T) part, 
but to deal with several large four-dimensional arrays, namely intermediate date, in the CCSD 
iteration~\cite{rendell:1992A,Kobayashi:1997A,anisimov:2014A,harding:2008A}. 
These $O(N^4)$ intermediate data often include electron repulsion integrals (ERIs),
intermediate results in the two-step tensor contractions, and some CCSD amplitude tensors
to construct new trial amplitudes (see section II for more details). For chemically interesting 
applications, these intermediate data can be terabyte-scale and thus cannot be stored in 
the available random access memory (RAM) on a single compute node. The traditional strategy of storing 
these data arrays to hard disks with heavy disk input/output (IO) traffic is obviously not an option 
for an efficient massive-parallel computation.

A data strategy that stores (part of) large arrays in the distributed RAM of compute nodes has been 
utilized in several CC codes in recent high-performance computing (HPC) packages, for example 
NWChem~\cite{nwchem:2011A, anisimov:2014A},Q-Chem~\cite{shao:2015A}, GAMESS~\cite{schmidt:1993A,olson:2007A}, 
Aquarius~\cite{solomonik:2014A} and MPQC~\cite{mpqc:2008A}, to name a few. In order to access the 
remote memory storage, the distributed-memory concept 
has to be fulfilled either in terms of the basic message passing interface (MPI) directly or 
employing sophisticated array distribution toolkits, like the global arrays (GA) shared memory
model~\cite{nieplocha:1996A} which has been used in NWChem~\cite{nwchem:2011A} and GAMESS~\cite{asadchev:2013A}. 
Furthermore, some other versatile toolkits proposed recently integrate advanced tensor utilities 
into self-defined distributed-memory array frameworks, which can significantly simplify the 
massive-parallel CC code but retaining high parallel efficiency. These toolkits include 
the so-called Cyclops Tensor Framework (CTF)~\cite{solomonik:2014A} used in 
Aquarius~\cite{solomonik:2014A} and Q-Chem packages~\cite{shao:2015A}, and the TiledArray tensor 
framework used in the MPQC package~\cite{mpqc:2008A}. 

The distributed-memory parallel closed-shell CCSD implementation using the GA toolkit in NWChem
was proposed in 1997 by Kobayashi and Rendell~\cite{nwchem:2011A}. In order to avoid the storing of
the whole ERI array in molecular orbital (MO) basis, Rendell's algorithm~\cite{rendell:1992A}
was adopted, \emph{i.e.}\ reformulating the contributions 
involving ERIs with three and four virtual (unoccupied) MO indices into the atomic-orbital (AO) 
basis and computing the relevant AO integrals on-the-fly, namely AO integral-direct algorithm.
A strong-scaling test on Cray T3D demonstrated excellent parallel efficiency (about 70\%) by 
increasing the number of processors from 16 to 256. 

Thereafter, Anisimov and co-workers realized that, when using more processors, the performance-limiting factor 
becomes the intensive network communication of doubles amplitudes in the tensor contraction 
of evaluating the so-called particle-particle ladder (pp-Ladder) term, which is the time-determining 
step in CCSD energy calculations~\cite{anisimov:2014A}. 
It has lead to a recent progress in NWChem which replicates the symmetrical doubles amplitudes
to reduce the communication in the AO integral-direct algorithm~\cite{anisimov:2014A}. 
The improved CCSD code in NWChem provides a notable speedup compared to the original 
distributed-memory code on 1,100 nodes and exhibits a good strong scaling from 1,100 to 20,000 nodes 
on ``Cray XE6'' supercomputer with 64 GB RAM per node~\cite{anisimov:2014A}. 
A similar replication algorithm had been used by Harding, \textit{et al.}\ to implement the CC methods 
in the MAB variant of the ACES II package~\cite{harding:2008A}.
However, the replication algorithm significantly increases the memory consumption and
somewhat weakens the memory scalability of the code.

To alleviate the distributed-memory communication latency, CTF~\cite{solomonik:2014A}
resorts to the use of the hybrid MPI/OpenMP communication layer
together with the sophisticated (so-called \emph{communication-optimal}) SUMMA algorithm 
in 2.5D variant~\cite{Geijn:1997A} for tensor contractions.
To take advantage of the CTF utility, the current CCSD implementations in Aquarius and Q-Chem 
are employing an open-shell CC formulation even for
closed-shell cases; while all intermediate arrays are fully 
stored and distributed in memory, including the most expensive one -- ERIs in MO basis. 
The CTF-based CCSD implementations present an excellent strong scaling in the Cray XC30 
supercomputer~\cite{solomonik:2014A}. 
For systems with hundreds of electrons and MOs, the parallel efficiency retains about 50\% 
with thousands of cores. 

The closed-shell CCSD implementation in MPQC was recently proposed by Peng \emph{et.\ al.} in 
2016~\cite{peng:2016A}. It uses the TiledArray toolkit to distribute in memory all necessary 
intermediate arrays with more than one index. TiledArray is an open-source framework for 
distributed-memory parallel implementation of dense and block-sparse tensor arithmetic,
which features a SUMMA-style communication algorithm with a task-based formulation.
Both conventional MO-only and AO integral-direct approaches are implemented. 
For the first approach, the resolution-of-identity (RI) approximation is taken to reduce 
the computational load of ERIs in MO basis. A distinctive feature of Peng's implementation 
is to completely turn off the use of permutation symmetry, particularly in the rate-limiting 
tensor contractions of evaluating the pp-Ladder term in CCSD equations.  Despite significantly 
increasing the computational cost in terms of floating point operations (FLOP), about three 
times more expensive than Rendell's algorithm, this choice allows for excellent parallel 
performance demonstrated to be scalable from a standalone multicore workstation with 16 cores to 
32 nodes on Blue-Ridge supercomputer with 408 nodes hosted by Virginia Tech Advanced Research 
Computing (VT ARC).

Beside aforementioned distributed-memory CC codes, there are many other CC algorithms implemented
in the same packages or others. To name a few in brief, another CC implementation in NWChem is
based on the Tensor Contraction Engine (TCE)~\cite{hirata:2003A,hu:2014A}; 
the original CC codes in Q-Chem use a general tensor contraction
library (so-called libtensor) for shared-memory architecture~\cite{epifanovsky:2013A,shao:2015A};
the parallel data strategies used in GAMESS include the Distributed Data Interface (DDI) 
for intra-node parallelism~\cite{olson:2007A} and the hybrid local disk+GA model~\cite{asadchev:2013A};
The ACES III package uses the super instruction assembly language (SIAL) for distributed memory
tensor contractions in CC theory~\cite{deumens:2011A};
and the closed-shell AO-driven CCSD and CCSD(T) methods in the PQS package using 
the Array Files (AF) scheme~\cite{ford:2007A}.
These CC implementations in quantum chemistry are all using Gaussian-type basis sets; 
but note that the CTF toolkit has been used very recently together with plane wave basis sets to 
provide CC-quality results for solid-state systems~\cite{Gruneis:2017A}.

In this paper, we describe a massive-parallel implementation of RI-based CCSD(T) for molecules
using the numeric atom-center orbital (NAO) basis sets in the Fritz Haber Institute \textit{ab initio}
molecular simulations package (FHI-aims)~\cite{blum:2009A}. A domain-based distributed-memory strategy 
upon the hybrid MPI/OpenMP communication layer is proposed
to replicate the intermediate data across domains, while distributing them among the computer nodes 
associated with the same domain. It allows for effectively utilizing the overall memory 
capacity to reduce the inter-domain communication latency without losing the parallel scalability 
for larger systems. Motivated by Peng's algorithm~\cite{peng:2016A}, we partially turn off the 
permutation symmetry in the rate-limiting tensor contraction steps, which reduces the difficulty
of designing a load-balanced distributed-memory strategy and alleviates the intra-domain 
communication latency. The sub-tensor contractions in each processor are carried out using 
multi-threaded BLAS library, and the data movements among processors are performed via unblocked 
MPI two-side communication scheme. In this first CCSD(T) implementation in FHI-aims, we do not 
equip our code with sophisticated tensor contraction toolkits, but this will be added soon. 
We expect a further improvement on the intra-domain tensor operations, in particular by 
employing CTF and/or TiledArray toolkits.

In Section II, we describe in detail the domain-based distributed-memory strategy in the context 
of evaluating the time-determining step in CCSD energy calculations, \textit{i.e.}\ the pp-Ladder 
term; and we discuss the key advantage of our algorithm of partially turning off the permutation 
symmetry in the same context. Section III documents the performance of our code on a single 
computer node against three state-of-the-art CCSD implementations with excellent shared-memory 
parallelism in the packages of MPQC, Psi4~\cite{Psi4}, and ORCA~\cite{ORCA}. 
We also benchmark the strong-scaling performance of our implementation with
up to 512 nodes with 10,000 cores in total and 128 GB RAM per node,
and compare with the distributed-memory CCSD codes implemented in Aquarius, MPQC, and NWChem.

FHI-aims provides a series of NAO basis sets with valence-correlation consistency, 
termed NAO-VCC-$n$Z with 
$n=2,3,4,5$, which ensures wave function-based methods consistently converging to 
the complete basis-set (CBS)
limit~\cite{zhang:2013A}. The RI implementation in FHI-aims features a prescription of 
producing an accurate, method-independent auxiliary basis set, which automatically adapts to 
a given basis set in a given system, therefore preventing a potential bias by the 
auxiliary basis sets optimized for other methods~\cite{ren:2012B,ihrig:2015A}. 

In Section IV, we demonstrate 1) our RI-CCSD(T) produces the exact CCSD(T) results for 
\emph{absolute total energies} using the same basis sets; 2) with the aid of the 
extrapolation scheme and the composite approach, our RI-CCSD(T) with NAO-VCC-$n$Z provides accurate 
results for several widely used quantum-chemistry molecular test sets.


\section{Theory and Implementation}
\label{section:theory-and-implementation}

The closed-shell CCSD implementation in this work takes the spin-adapted formulation of Scuseria,
Janssen, and Schaefer~\cite{scuseria:1988A}, which is accelerated by the direct inversion of 
the iterative subspace (DIIS) method~\cite{scuseria:1986A}.
Using the converged CCSD amplitudes, the non-iterative evaluation of perturbative triples energy is
then implemented based on the algorithm of Rendell and Lee~\cite{rendell:1991A,rendell:1993A}.
Instead of repeating the description of the well-documented CC theory and CCSD(T) formalisms, 
we will give details only for the key modifications employed in our approach.

As in the usual convention, we will use symbols $i,j,k,l$ and $N_{occ}$ to denote the indices and 
the number of occupied MOs; and $a,b,c,d$ and $N_{vir}$ for unoccupied MOs. Meanwhile,
$p,q,r,s$ and $N_{MO}$ will be used to label the indices and the number of unspecified MOs; 
and letters of Greek alphabet $\alpha,\beta,\gamma,\eta,\mu,\nu$ with $N_{(A)BS}$ for the (auxiliary) 
AO basis sets. The index values all start from 1.
We focus on the massive-parallel implementation of closed-shell CCSD(T), 
therefore the indices are spin-free throughout the paper.
Note that, the open-shell version of CCSD has been coded in FHI-aims as well 
but not yet fully optimized; and the open-shell perterbative (T) implementation is 
not yet available.

\subsection{Formalism}
\label{section:theory}
The CCSD wave function is obtained from the Hartree-Fock single slater determinant 
ground state 
$\vert\Psi_{0}\rangle$:
\begin{equation}
 \vert\Psi_{CCSD}\rangle=e^{\hat{T}}\vert\Psi_{0}\rangle,
\end{equation}
with the exponential cluster operator $\hat{T}$
\begin{equation}
 \hat{T}=\hat{T}_1+\hat{T}_2.
\end{equation}
$\hat{T}_1$ and $\hat{T}_2$ are spin-free single and double excitation operators for 
closed-shell systems, which can be described in the unitary group approach,
\begin{equation}
 \label{eq:t1}
 \hat{T}_1=\sum_{ia} t_{i}^{a}\hat{E}_{i}^{a},
\end{equation}
\begin{equation}
 \label{eq:t2}
 \hat{T}_2=\frac{1}{2}\sum_{ijab} t_{ij}^{ab}\hat{E}_{i}^{a}\hat{E}_{j}^{b},
\end{equation}
with the definition of the unitary group generator $\hat{E}_{q}^{p}$ as
\begin{equation}
 \hat{E}_{q}^{p}=\hat{a}_{p}^{\dagger}\left(\alpha\right)\hat{a}_{q}\left(\alpha\right)
 +\hat{a}_{p}^{\dagger}\left(\beta\right)\hat{a}_{q}\left(\beta\right).
\end{equation}
Here $\hat{a}$ and $\hat{a}^{\dagger}$ are annihilation and creation operators, 
while $\alpha$ and $\beta$ refer to the spin states.
The amplitudes of single- and double-exciation configurations are $t_{i}^{a}$ (or $t_{s}$) and 
$t_{ij}^{ab}$ (or $t_{d})$. In our approach, all arrays with only one or two
indieces, including $t_{i}^{a}$, will be replicated in each processor. $t_{i}^{a}$ in
double precision consumes only $8N_{occ}N_{vir}$ bytes; while $t_{ij}^{ab}$ is allocated with
double prcision and without symmetry, thus consuming $8N_{occ}^2N_{vir}^2$ bytes.

$t_{i}^{a}$ and $t_{ij}^{ab}$ are determined by solving the CCSD equation,
which project the Schr\"{o}dinger equation onto the Hartree-Fock ground state $\vert\Psi_{0}\rangle$, 
single-excitation states $\vert\Psi_{i}^{a}\rangle$, and double-excitation states $\vert\Psi_{ij}^{ab}\rangle$ as follow,
\begin{equation}
	\label{eq:ccsd-1}
	\langle\Psi_{0}\vert e^{-\hat{T}} \left(\hat{H}-E_{0} \right)e^{\hat{T}}\vert \Psi_{0}\rangle=E_{corr}^{\textrm{CCSD}},
\end{equation}

\begin{equation}
	\label{eq:ccsd-2}
 \langle\Psi_{i}^{a}\vert e^{-\hat{T}} \left(\hat{H}-E_{0} \right)e^{\hat{T}}\vert \Psi_{0}\rangle=0,
\end{equation}

\begin{equation}
	\label{eq:ccsd-3}
 \langle\Psi_{ij}^{ab}\vert e^{-\hat{T}} \left(\hat{H}-E_{0} \right)e^{\hat{T}}\vert \Psi_{0}\rangle=0.
\end{equation}
Here $\hat{H}$ is the Hamiltonian for real systems, $E_{0}$ the Hartree-Fock ground-state energy, and 
$E_{corr}^{\textrm{CCSD}}$ the CCSD correlation energy.
The CCSD equations have to be solved iteratively if following the most widely used Jacobi 
solver~\cite{scuseria:1986A}. With the converged $t_{i}^{a}$ and $t_{ij}^{ab}$, 
the closed-shell CCSD and perturbative (T) correlations can be evaluated from the following equations:
\begin{equation}
	E_{corr}^{\textrm{CCSD}}=\sum_{ijab}\left( 2v_{ab}^{ij}-v_{ba}^{ij}\right) (t_{ij}^{ab}+t_{i}^{a}t_{j}^{b}),
\end{equation}
\begin{equation}
	E_{corr}^{\textrm{(T)}}=\sum_{ijk}\sum_{abc}\frac{\left(4W_{ijk}^{abc}+W_{ijk}^{bca}+W_{ijk}^{cab}\right)
	\left(V_{ijk}^{abc}-V_{ijk}^{cba}\right)}{3D_{ijk}^{abc}}.
\end{equation}
In this (T) energy expression,
\begin{equation}
	W_{ijk}^{abc}=P_{ijk}^{abc}\left(\sum_{d}^{vir}v_{di}^{ba}t_{kj}^{cd}-\sum_{l}^{occ}v_{kl}^{cj}t_{il}^{ab}\right),
\end{equation}
\begin{equation}
	V_{ijk}^{abc}=W_{ijk}^{abc}+v_{jk}^{bc}t_{i}^{a}+v_{ik}^{ac}t_{j}^{b}+v_{ij}^{ab}t_{k}^{c},
\end{equation}
and
\begin{equation}
	D_{ijk}^{abc}=\epsilon_{i}+\epsilon_{j}+\epsilon_{k}-\epsilon_{a}-\epsilon_{b}-\epsilon_{c},
\end{equation}
with $v_{ab}^{ij}$ being the electron repulsion integrals of molecular orbitals 
(see the next subsection for more details).
$P_{ijk}^{abc}$ is the permutation operator $P_{ijk}^{abc}f(ijk,abc)=f(ijk,abc)+
f(ikj,acb)+f(kij,cab)+f(kji,cba)+f(jki,bca)+f(jik,bac)$, and $\epsilon_{p}$ is the Hartree-Fock
eigenvalue of molecular orbital $p$.

\subsection{Intermediate data}
\label{section:intermediate}
As shown above, in addtion to the single and double amplitudes $t_{i}^{a}$ and $t_{ij}^{ab}$, 
several large arrays with four indices, namely intermediate data, are retrieved (and/or updated) 
heavily during the CCSD iteration, which mainly include
\begin{itemize}
	\item $t_{d}^{(n)}$: doubles amplitudes in the $n$th iteration step.
		In order to accelerate the CCSD iteration using the DIIS strategy, several doubles amplitudes in 
		previous steps should be recorded. (Often 3 steps are enough in practical use.)
		In our approach, the symmetry is used to reduce the storage of $t_{d}^{(n)}$ to
		$4N_{occ}N_{vir}(N_{occ}N_{vir}+1)$ bytes in double precision.
	\item $v_{rs}^{pq}$: electron repulsion integrals of molecular orbitals 
		$\{\phi_p(\boldsymbol{r})\}$, namely ERIs in MO basis or MO-ERIs: 
        \begin{equation}
         \label{eq:eri}
         v_{rs}^{pq}=
                \int\frac{\phi_p^*(\boldsymbol{r}_1)\phi_r(\boldsymbol{r}_1)\phi_q^*(\boldsymbol{r}_2)\phi_s(\boldsymbol{r}_2)}
                    {\left|\boldsymbol{r}_1-\boldsymbol{r}_2\right|}
                 d\boldsymbol{r}_1d\boldsymbol{r}_2,
        \end{equation}
		which consume $8N_{MO}^4$ bytes in double precision without symmetry.
\end{itemize}
These intermediate arrays share the same $O(N^4)$ scaling in memory consumption, but it is obvious that
the storage of $v_{rs}^{pq}$ is the most challenging one, because $N_{MO}=N_{occ}+N_{vir}$ and 
$N_{vir}\gg N_{occ}$ in practical use. Despite the full $v_{rs}^{pq}$ tensor should be used in CCSD(T)
calculations,
it is convenient to handle them separately in terms of the number of unoccupied indices,
resulting in several sub-tensors as $v_{kl}^{ij}$, $v_{ka}^{ij}$, $v_{ab}^{ij}$, $v_{bc}^{ia}$, and $v_{cd}^{ab}$. 
Apparently, the MO-ERIs with four unoccupied indices $v_{cd}^{ab}$ remain to be the most
memory-demanding part. More importantly, as shown in the following discussion, 
$v_{cd}^{ab}$ is involved in the evaluation of pp-Ladder contribution 
(see equation~\ref{eq:pp-Ladder-0}), making the manipulation of $v_{cd}^{ab}$ crucial
for massive parallelism.
Note that MOs $\{\phi_p(\boldsymbol{r})\}$ are expanded in terms of a set of AO basis functions 
$\{\psi_\alpha(\boldsymbol{r})\}$
\begin{equation}
	\phi_p=\sum_{\alpha=1}^{N_{BS}}c_{p\alpha}\psi_\alpha(\boldsymbol{r}),
\end{equation}
with the expension coefficients as $\{c_{p\alpha}\}$.
The AO integral-direct algorithm proposed by Rendell avoids the direct storage of 
$v_{cd}^{ab}$, which reformulates the contribution involving the use of $v_{cd}^{ab}$ in the 
AO representation and computes the relevant AO-ERIs ($v_{\alpha\beta}^{\gamma\eta}$)
on-the-fly~\cite{rendell:1992A,Kobayashi:1997A}.
The AO integral-direct algorithm is mainly applied with Gaussian-type basis 
sets~\cite{peng:2016A,scuseria:1988A,saebo:1987A,Kobayashi:1997A,janowski:2007A}.

Resolution of identity (RI) approximation (also known as ``density fitting (DF)'') is now the most 
successful approach to alleviate the computational load of ERIs\cite{ren:2012A,zhang:2015A}. 
The RI approach allows for the decomposition of the fourth-rank ERI tensor in terms of third-rank tensors, 
and is therefore better suited to be pre-stored:
\begin{equation}
\label{eq:RIM}
 v_{rs}^{pq}\approx\sum_{\mu}m_{pr}^{\mu}m_{qs}^{\mu},
\end{equation}
where the index $\mu$ runs over an auxiliary basis $\{P_{\mu}(\boldsymbol{r})\}$ with the size of $N_{ABS}$.
\begin{itemize}
	\item $m_{pr}^{\mu}$: the decomposed third-rank tensor in MO basis, which consumes $8N_{MO}^2N_{ABS}$ bytes
		in double precision. In RI-V approximation, $m_{pr}^{\mu}$ is determined by directly
		minimizing the errors in the AO-ERIs \cite{ren:2012A}:
    	\begin{equation}
    	 \label{eq:ri-v}
    	 m_{pr}^{\mu}=\sum_{\nu,\alpha\beta}\left(\alpha\beta|\nu\right)\left(\mu|\nu\right)^{-1/2}c_{p\alpha}^*c_{r\beta}.
    	\end{equation}
		Here $\left(\alpha\beta|\nu\right)$ is the three-center integral between the AO basis pair 
		$\psi_\alpha^*\psi_\beta$ and the auxiliary basis $P_{\nu}$
		\begin{equation}
		 \left(\alpha\beta|\nu\right)=\int \frac{\psi_\alpha^*(\boldsymbol{r}_1)\psi_\beta(\boldsymbol{r}_1)P_\nu(\boldsymbol{r}_2)}
			 {\left|\boldsymbol{r}_1-\boldsymbol{r}_2\right|}
		         d\boldsymbol{r}_1d\boldsymbol{r}_2,
		\end{equation}
		and $\left(\mu|\nu\right)$ is the two-center Coulomb integral for the auxiliary basis functions,
		\begin{equation}
		 \left(\mu|\nu\right)=\int \frac{P_{\mu}^*(\boldsymbol{r}_1)P_\nu(\boldsymbol{r}_2)}
			 {\left|\boldsymbol{r}_1-\boldsymbol{r}_2\right|}
		         d\boldsymbol{r}_1d\boldsymbol{r}_2.
		\end{equation}
\end{itemize}
RI methods and other tensor decomposition techniques, like the partial Cholesky 
decomposition(CD)~\cite{beebe:1977A,roeggen:1986A} and the tensor hypercontraction 
scheme~\cite{hohenstein:2012A}, have been applied to the CCSD and/or CCSD(T) 
implementation~\cite{rendell:1994A,head-gordon:2010A,
pitonak:2011A,hohenstein:2012B,benedikt:2013A,deprince:2013A,parrish:2014A,peng:2016A}. 
DePrince and co-workers demonstrated that the auxiliary basis sets optimized for MP2 theory are able 
to provide accurate RI-CCSD(T) results for weak interactions and reaction energies~\cite{deprince:2013A}.
Furthermore, it has been proposed that tensor decomposition techniques can be used to either reduce 
the scaling of the most vexing term in CCSD equations \cite{parrish:2014A} or even the whole CCSD 
equations \cite{head-gordon:2010A,benedikt:2013A,hohenstein:2012B}. 
In our approach with NAO basis sets, $v_{cd}^{ab}$ will be calculated on-the-fly in the RI-V 
fashion during CCSD(T) calculations; while the rest MO-ERIs with fewer unoccupied indices will be 
pre-stored and distributed in memory directly

\begin{table}[!h]
	\begin{ruledtabular}
\caption{Memory requirement of several intermediate arrays for a set of molecules 
         investigated in this paper with different system sizes (unit GB).}
\label{table:memory}
\begin{tabular}{ccccccc}
	Complex\footnote{The basis sets used in these calculations are cc-pVDZ for water clusters and 
	a modified triple-zeta basis set (mTZ) for $\beta$-carotene~\cite{hu:2014A,peng:2016A}.}
	& $N_{occ}$ & $N_{vir}$ & $N_{ABS}$ & $m_{pr}^{\mu}$ & $t_{ij}^{ab}$ & $v_{bc}^{ia}$\\
\hline
(H$_2$O)$_{10}$ &  50 & 190 & 920  &  0.42 &  0.72 &   2.74 \\
(H$_2$O)$_{15}$ &  75 & 285 & 1380 &  1.43 &  3.66 &  13.89 \\
(H$_2$O)$_{20}$ & 100 & 380 & 1840 &  3.39 & 11.55 &  43.90 \\
$\beta$-carotene& 108 & 884 & 4504 & 35.46 & 72.92 & 596.86 \\
\end{tabular}
\end{ruledtabular}
\end{table}

\begin{itemize}
	\item $v_{ij}^{kl}$, $v_{ij}^{ak}$, and $v_{ij}^{ab}$: MO-ERIs with zero, one, and two 
		unoccupied indices,which consume $8N_{occ}^4$ bytes, $8N_{occ}^{3}N_{vir}$ bytes, and
		$8N_{occ}^2N_{vir}^{2}$ bytes, respectively in double precision.
	\item $v_{bc}^{ia}$: MO-ERIs with three unoccupied indices which consumes $8N_{occ}N_{vir}^{3}$ bytes
		in memory.
		In order to obtain the good performance for the perturbative triple (T) evaluation, 
		our approach follows the 
		idea of Rendell, Lee, and Komornicki~\cite{rendell:1991A} to pre-store and
		distribute $v_{bc}^{ia}$ in memory.
		But for the tensor contractions with $v_{bc}^{ia}$ which are not locally pre-stored, 
		RI-V approximation will be used to reduce the communication. 
\end{itemize}

In consequence, our approach will allocate and/or prepare a series of forth-rank tensors with 
one third-rank RI-V tensor before the CCSD iteration procedure. Table~\ref{table:memory} 
provides the memory consumption of these intermediate data for a set of molecules. 
Not surprisingly, the memory requirement increases dramatically in terms of the system size. 
Taking the cluster with 20 water molecules (H$_2$O)$_{20}$ with the cc-pVDZ 
basis set as example, it consumes 11.5 GB for the doubles amplitudes, 3.4 GB 
for $m_{pr}^{\mu}$, and 43.9 GB for $v_{bc}^{ia}$. 
Considering that several temporary buffer files with a similar size of doubles amplitudes 
are needed (see section~\ref{section:dbdm} for details), 
the memory requirement in total largely exceeds the memory
capacity of a single computer node in today's supercomputers. Thus the data should be distributed 
over many computer nodes to fully utilize the global memory capacity.

\subsection{The pp-Ladder evaluations}
\label{section:pp-ladder}
As widely discussed in the literature~\cite{rendell:1992A,Kobayashi:1997A,anisimov:2014A,
harding:2008A}, massive-parallel CCSD(T) calculations are essentially communication bound,
particularly in calculating the pp-Ladder diagram generated
in equation~\ref{eq:ccsd-3}. The resulting pp-Ladder array $L_{ij}^{ab}$ shares the same size of
doubles amplitudes and is contracted by
\begin{equation}
	\label{eq:pp-Ladder-0}
	L_{ij}^{ab}=\sum_{cd}b_{cd}^{ab}\tau_{ij}^{cd},
\end{equation}
with the definitions of $\tau_{ij}^{ab}$ as
\begin{equation}
	\label{eq:pp-Ladder-01}
 \tau_{ij}^{ab}=t_{ij}^{ab}+t_{i}^{a}t_{j}^{b},
\end{equation}
and $b_{cd}^{ab}$ (as large as $v_{cd}^{ab}$) as 
\begin{equation}
	\label{eq:pp-Ladder-1}
	b_{cd}^{ab}=v_{cd}^{ab}-\sum_{k}v_{cd}^{ak}t_{k}^{b}-\sum_{k}t_{k}^{a}v_{cd}^{kb}.
\end{equation}
In our approach, the RI-V approximation is used to evaluate $v_{cd}^{ab}$ and $v_{cd}^{ak}$ on-the-fly, 
leading to the construction of $b_{cd}^{ab}$ as
\begin{equation}
	\label{eq:pp-Ladder-2}
	b_{cd}^{ab}=\sum_{\mu}m_{ac}^{\mu}\left(m_{bd}^{\mu}-\bar{m}_{bd}^{\mu}\right)-\sum_{k}t_{k}^{a}v_{cd}^{kb},
\end{equation}
with
\begin{equation}
	\label{eq:pp-Ladder-21}
	\bar{m}_{bd}^{\mu}=\sum_{k}m_{kd}^{\mu}t_{k}^{b}.
\end{equation}
The hybrid-RI algorithm for MO-ERIs with three unoccupied indices is designed to balance the 
computing cost and communication in our
domain-based distributed-memory strategy which will be introduced in the following section. 
As a result, the total cost of evaluating the pp-Ladder array in terms of 
FLOP is $N_{vir}^4(N_{occ}^2+2N_{ABS}+N_{occ}+N_{occ}N_{ABC}N_{vir}^{-2})$, in which
the leading term is the tensor contraction of equation~\ref{eq:pp-Ladder-0} with $O(N^6)$ scaling.

In the seminal paper of CCSD formulation~\cite{scuseria:1988A}, Scuseria \textit{et al.}\ 
innovatively divided $L_{ij}^{ab}$ into two auxiliary, but highly symmetric tensors, 
which reduce the FLOP count of the leading term from $N_{occ}^2N_{vir}^4$ to 
$\frac{1}{4}N_{occ}^2N_{vir}^4$.
Scuseria's formulation provides the optimal computational cost for CCSD and has been widely used in 
quantum-chemistry codes, such as Gaussian~\cite{g16}, ORCA~\cite{ORCA}, \emph{etc}. 
However, in line with the observation of previous works on massiv-parallel CC methods, 
we find that FLOP count is not the only issue relevant to the computational cost. Other aspects 
as the I/O operation, communication, load-balance, and vectorization 
are also of (often vital) importance.

Our implementation only takes advantage of the symmetry in equation~\ref{eq:pp-Ladder-0}
\begin{equation}
	L_{ij}^{ab}=L_{ji}^{ba}\textrm{ and }\tau_{ij}^{ab}=\tau_{ji}^{ba}. 
\end{equation}
This reduces the numerical workload to double loops with $i\le j$. Accordingly, we introduce 
a new index $\lambda$ with $\lambda=j(j-1)/2+i$, and note the relevant tensors as $L_{\lambda}^{ab}$ 
and $\tau_{\lambda}^{ab}$. In addition to the advantage of reducing the leading cost 
(equation~\ref{eq:pp-Ladder-0}) to $\frac{1}{2}N_{occ}^2N_{vir}^4$,
this approach does not introduce additional manipulation of $b_{cd}^{ab}$ -- the largest intermediate 
tensor in the work, making it easier to design a load-balanced distributed-memory strategy 
with reduced communication. 

\subsection{Domain-based distributed-memory strategy}
\label{section:dbdm}

It is well-documented that, for massive-parallel calculations, the CCSD implementation based 
on the standard distributed-memory strategy is encountering the so-called communication bottleneck.
It is because the large intermediate data mentioned above are distributed globally and uniformly 
over the processors. The data communication becomes unaffordable very quickly with the 
increasing number of processors. 
For convenience, we use symbols $x$ and $N_p$ to denote the index and the number of processors 
$\{P(x)\}$ in the standard distributed-memory strategy.

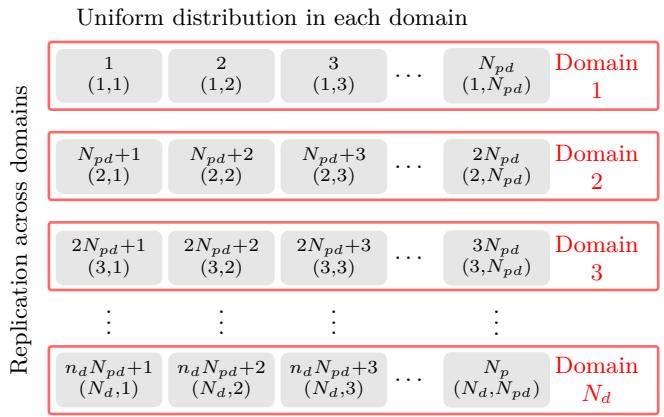
\begin{figure}
	\begin{tikzpicture}[remember picture]
		\node[format, text width=1.2cm, align=center] (d00) {
			1\\
			(1,1)
		};
		\node[format, text width=1.2cm, align=center, right = 0.06cm of d00] (d01) {
			2\\
			(1,2)
		};
		\node[format, text width=1.2cm, align=center, right = 0.06cm of d01] (d02) {
			3\\
			(1,3)
		};
		\node[text width=0.5cm, align=center, right = -0.06cm of d02] (d03) {
			$\dots$
		};
		\node[format, text width=1.2cm, align=center, right = 0.06cm of d03] (d04) {
			$N_{pd}$\\
			(1,$N_{pd}$)
		};
		\node[text width=1.2cm, color=red, align=center, right = -0.10cm of d04] (d05) {
			Domain 1
		};
		\draw[color=red!60, line width = 1pt, rounded corners=1pt] 
		(-0.8cm,0.45cm) -- (-0.8cm,-0.45cm) -- (7.3cm,-0.45cm) -- (7.3cm,0.45cm) -- cycle;

		\node[format, text width=1.2cm, align=center, below = 0.47cm of d00] (d10) {
			$N_{pd}$+1\\
			(2,1)
		};
		\node[format, text width=1.2cm, align=center, right = 0.06cm of d10] (d11) {
			$N_{pd}$+2\\
			(2,2)
		};
		\node[format, text width=1.2cm, align=center, right = 0.06cm of d11] (d12) {
			$N_{pd}$+3\\
			(2,3)
		};
		\node[text width=0.5cm, align=center, right = -0.06cm of d12] (d13) {
			$\dots$
		};
		\node[format, text width=1.2cm, align=center, right = 0.06cm of d13] (d14) {
			$2N_{pd}$\\
			(2,$N_{pd}$)
		};
		\node[text width=1.2cm, color=red, align=center, right = -0.10cm of d14] (d15) {
			Domain 2
		};
		\draw[color=red!60, line width = 1pt, rounded corners=1pt] 
		(-0.8cm,-0.75cm) -- (-0.8cm,-1.65cm) -- (7.3cm,-1.65cm) -- (7.3cm,-0.75cm) -- cycle;

		\node[format, text width=1.2cm, align=center, below = 0.47cm of d10] (d20) {
			$2N_{pd}$+1\\
			(3,1)
		};
		\node[format, text width=1.2cm, align=center, right = 0.06cm of d20] (d21) {
			$2N_{pd}$+2\\
			(3,2)
		};
		\node[format, text width=1.2cm, align=center, right = 0.06cm of d21] (d22) {
			$2N_{pd}$+3\\
			(3,3)
		};
		\node[text width=0.5cm, align=center, right = -0.06cm of d22] (d23) {
			$\dots$
		};
		\node[format, text width=1.2cm, align=center, right = 0.06cm of d23] (d24) {
			$3N_{pd}$\\
			(3,$N_{pd}$)
		};
		\node[text width=1.2cm, color=red, align=center, right = -0.10cm of d24] (d25) {
			Domain 3
		};
		\draw[color=red!60, line width = 1pt, rounded corners=1pt] 
		(-0.8cm,-1.95cm) -- (-0.8cm,-2.85cm) -- (7.3cm,-2.85cm) -- (7.3cm,-1.95cm) -- cycle;

		\node[text width=1.2cm, align=center, below = 0.0cm of d20] (d30) {
			$\vdots$
		};
		\node[text width=1.2cm, align=center, below = 0.0cm of d21] (d31) {
			$\vdots$
		};
		\node[text width=1.2cm, align=center, below = 0.0cm of d22] (d32) {
			$\vdots$
		};
		\node[text width=1.2cm, align=center, below = 0.0cm of d24] (d34) {
			$\vdots$
		};

		\node[format, text width=1.2cm, align=center, below = 0.12cm of d30] (d40) {
			$n_{d}N_{pd}$+1\\
			($N_{d}$,1)
		};
		\node[format, text width=1.2cm, align=center, right = 0.06cm of d40] (d41) {
			$n_{d}N_{pd}$+2\\
			($N_{d}$,2)
		};
		\node[format, text width=1.2cm, align=center, right = 0.06cm of d41] (d42) {
			$n_{d}N_{pd}$+3\\
			($N_{d}$,3)
		};
		\node[text width=0.5cm, align=center, right = -0.06cm of d42] (d43) {
			$\dots$
		};
		\node[format, text width=1.2cm, align=center, right = 0.06cm of d43] (d44) {
			$N_{p}$\\
			($N_{d}$,$N_{pd}$)
		};
		\node[text width=1.2cm, color=red, align=center, right = -0.10cm of d44] (d45) {
			Domain $N_{d}$
		};
		\draw[color=red!60, line width = 1pt, rounded corners=1pt] 
		(-0.8cm,-3.60cm) -- (-0.8cm,-4.50cm) -- (7.3cm,-4.50cm) -- (7.3cm,-3.60cm) -- cycle;

		\node[rotate=90, text width=4.6cm, align=center, above left = 0.0cm and 0.2cm of d00] {
			Replication across domains 
		};
		\node[text width=6.0cm, align=center, above left = 0.2cm and -6.0cm of d00] {
			Uniform distribution in each domain 
		};
	\end{tikzpicture}
	\caption{\label{fig:dbdm} Illustration of the domain-based distributed-memory strategy 
		used in our approach to replicate and distribute intermediate data across processors.
		For short, we define $n_d=N_{d}-1$ to index processors in the last domain.}
\end{figure}

To alleviate this problem, a domain-based distributed-memory strategy is proposed, 
which groups the processors to different domains. Specifically, we introduce an index $y$ to label
the domain of a certain processor. It results in a 2D-grid distribution
of compute processors $\{P(y,x)\}$ with $x$ labeling the processors in each domain.
In the current version, we restrict the number of processors per domain $N_{pd}$ to be the same,
\begin{equation}
	N_{p}=N_{d}*N_{pd}
\end{equation}
with the integer $N_{d}=1,2,3,\cdots$ denoting the number of domains. 
The system-dependent $N_{pd}$ is determined to ensure that the global memory of each domain
is sufficiently high to accommodate all the pre-stored intermediate data together with a certain amount of buffer space.
Figure~\ref{fig:dbdm} visualizes the 2D hierarchical organization of compute processors
in the domain-based distributed-memory strategy.
Since each domain now possesses a full copy of pre-stored intermediate data, it is very easy to design
the corresponding CC algorithm with negligible inter-domain communication load, thus by definition 
ensuring the scalability in massive parallelism. On the other hand, the double amplitudes $t_{ij}^{ab}$
and all intermediate data with more than two unoccupied or full MO indices will be uniformly 
distributed across the compute nodes in each domain.
Therefore, our domain-based distributed-memory strategy can be considered as a straightforward 
generalization of the standard distributed-memory strategy with optimal flexibility to fully utilize 
the global memory capacity and alleviate the communication latency at scale. By setting $N_{d}=1$, 
it will decay to the standard distributed-memory strategy.

In order to take advantage of the hybrid MPI/OpenMP model, it would be better to create 
fewer processors per node. Assuming that each node generates one processor, 
the 2D-grid hierarchy of processors shown in figure~\ref{fig:dbdm} 
is equivalent to that of compute nodes.
Note that, for supercomputers composed by dual-socket x86-based blades, 
optimal is one processor per socket (\textit{i.e.}\ two processors per node).

Our domain-based distributed-memory strategy requires a specific parallel algorithm. 
Let us take the evaluation of the pp-Ladder array $L_{\lambda}^{ab}$. 
The corresponding pseudocode is presented in figure~\ref{fig:pseudocode}. 
The important points are as follows:

\begin{itemize}
	\item The tensor contraction for $L_{\lambda}^{ab}$ is divided into $N_{d}$ sub-tensor 
		contractions labeled by the unoccupied index $a$. Since each domain possesses a 
		full copy of all pre-stored intermediate data, the resulting inter-domain sub-tensors 
		$\{(L_{\lambda}^{\bar{a}b})_{y}\}$ will be contracted in each domain $\left\{ y \right\}$ 
		without any inter-domain communication. $\bar{a}$ is the local unoccupied index in the 
		$y$th domain with $a=(\bar{a}-1)N_{d}+y$. 
        \begin{equation}
        	\label{eq:pp-Ladder-3}
			L_{\lambda}^{\bar{a}b}=\sum_{cd}b_{cd}^{\bar{a}b}\tau_{\lambda}^{cd}.
        \end{equation}
		In consequence, a series of sub-tensors $\{(b_{cd}^{\bar{a}b})_{y}\}$, with a similar
		inter-domain distribution is needed.

	\item In each domain, $L_{\lambda}^{\bar{a}b}$ is distributed 
		across compute nodes in terms of the index $\lambda$,
		resulting in the intra-domain sub-tensors $\{(L_{\underline\lambda}^{\bar{a}b})_{x}\}$.
		Here we define a new intra-domain local index $\underline\lambda$ in the $x$-th processor
		which obeys $\lambda=(\underline\lambda-1)N_{pd}+x$. 
		Meanwhile $b_{cd}^{\bar{a}b}$ and $\tau_{\lambda}^{cd}$ are uniformly distributed
		in each domain as $\{(b_{c\underline{d}}^{\bar{a}b})_x\}$ and 
		$\{(\tau_{\lambda}^{c\underline{d}})_{x}\}$.
		The newly introduced local index $\underline{d}$ indicates the distribution of 
		the unoccupied index $d$ with $d=(\underline{d}-1)N_{pd}+x$.  
		
	\item To minimize the intra-domain communication, pre-stored intermediate data are either 
		distributed before CCSD(T) calculations or re-distributed before evaluating the pp-Ladder 
		term:
		(A) $t_{ij}^{cd}\rightarrow \{ (t_{ij}^{c\underline{d}})_{x}\}$;
		(B) $m_{pd}^{\mu}\rightarrow \{ (m_{p\underline d}^{\mu})_{x}\}$;
		(C) $v_{cd}^{kb}\rightarrow \{(v_{c\underline{d}}^{kb})_{x}\}$.

	\item For the evaluation of $\{(v_{c\underline{d}}^{kb})_{x}\}$ a hybrid-RI strategy is employed.
		As the MO-ERIs are distributed with
		three indices in terms of $d$ (same as for $b_{cd}^{ab}$), 
		no intra-domain communication is needed in the last contraction of equation~\ref{eq:pp-Ladder-1}.
		However, for the second contraction of equation~\ref{eq:pp-Ladder-1}, this kind of 
		distribution cannot avoid intensive communication. This motivates the 
		hybrid-RI strategy, \textit{i.e.}\ using RI-V approximation for those $v_{cd}^{kb}$ that 
		are not pre-stored locally (see equation~\ref{eq:pp-Ladder-2}).
		
	\item Similar to $\{(v_{c\underline{d}}^{kb})_{x}\}$, the intra-domain distribution of the 
		RI-V tensor $\{ (m_{p\underline d}^{\mu})_{x}\}$ cannot avoid the communication to prepare
		$m_{\bar{a}c}^{\mu} \leftarrow \{(m_{\bar{a}\underline c}^{\mu})_{x}\}$, which are requested
		for all processors within domain to calculate $v_{cd}^{\bar{a}b}$ on-the-fly 
		(see equation~\ref{eq:pp-Ladder-1} and~\ref{eq:pp-Ladder-2}).
		Profiting by the domain-based concept, this intra-domain communication load decreases
		with $N_d$, because the workload with the loop of $a$ has been well parallelized
		across domains.

	\item Once all intermediate data have been evaluated, the evaluation of
		$L_{\lambda}^{\bar{a}b}$ will be accomplished by three steps: (1) divide
		the task into $N_{pd}$ sub-tasks $\{(L_{\underline\lambda}^{\bar{a}b})_{x}\}$, each of which
		takes over a block of $\{\lambda\}$, \textit{i.e.}\ $\{(\underline\lambda)_{x}\}$,
		and will be finally stored in processor $x$;
		(2) for each sub-task, calculate $(\underline{L}_{\underline\lambda}^{\bar{a}b})_{x}$
		\begin{equation}
			\begin{split}
				(\underline{L}_{\underline\lambda}^{\bar{a}b})_{x}
				=&\sum_{\underline{d}}\sum_{c}b_{c\underline{d}}^{\bar{a}b}
				(\tau_{\underline\lambda}^{c\underline{d}})_{x},
			\end{split}
		\end{equation}
		which is an incomplete contraction looping over local $\{\underline{d}\}$;
		(3) perform intra-domain communication to sum over 
		$\{(\underline{L}_{\underline\lambda}^{\bar{a}b})_{x}\}\rightarrow 
		(L_{\underline\lambda}^{\bar{a}b})_{x}$ and store in processor $x$.
		Unblocked MPI two-side communication scheme is utilized to balance the contraction 
		workload at step (2) and the communication latency at step (3).
\end{itemize}

\begin{figure}
	\begin{tikzpicture}[remember picture]
		\node[pseudo, text width=9.0cm, align=left] (d00) {
			Subroutine \texttt{CC\_cl\_add\_b2w}:
		};
		\node[pseudo, text width=8.5cm, align=left, below=0.0cm of d00] (d01) {
			do $\bar{a}$ = 1, $\bar{N}_{vir}$  (loop $a$, parallelized across domains) 
		};
		\node[pseudo, text width=8.0cm, align=left, below=0.0cm of d01] (d02) {
		    Load $m_{\bar{a}c}^{\mu}$ via intra-domain communication.\\ 
			Prepare $\tau_{\lambda}^{c\underline{d}}$ without communication.
		};
		\node[pseudo, text width=8.0cm, align=left, below=0.0cm of d02] (d03) {
			Prepare $b_{c\underline{d}}^{\bar{a}b}$ by 3 \texttt{dgemm}s
			using multi-threaded Intel BLAS library without communication.
		};
		\node[pseudo, text width=8.0cm, align=left, below=0.0cm of d03] (d10) {
			do $x$ = 1, $N_{pd}$  (loop $\lambda$ block by block with $\{(\underline{\lambda})_{x}\}$)
		};
		\node[pseudo, text width=7.5cm, align=left, below=0.0cm of d10] (d13) {
			Calculate $(\underline{L}_{\underline{\lambda}}^{\bar{a}b})_{x}$ by 
			1 multi-threaded \texttt{dgemm}.
		};
		\node[pseudo, text width=7.5cm, align=left, below=0.0cm of d13] (d14) {
			Sum over $(\underline{L}_{\underline\lambda}^{\bar{a}b})_{x}$ via
			intra-domain communication, and store the resulting $(L_{\underline\lambda}^{\bar{a}b})_{x}$ 
			in processor $x$.
		};
		\node[pseudo, text width=8.0cm, align=left, below=0.0cm of d14] (d10) {
			end do $x$
		};
		\node[pseudo, text width=8.5cm, align=left, below = 0.0cm of d10] (d00) {
			end do $\bar{a}$
		};
	\end{tikzpicture}
	\caption{\label{fig:pseudocode} Pseudocode of evaluating the pp-Ladder array $L_{ij}^{ab}$, 
		equations~\ref{eq:pp-Ladder-0}--\ref{eq:pp-Ladder-21}.}
\end{figure}

By partially turning off the use of symmetry in equation~\ref{eq:pp-Ladder-0} and employing 
the hybrid-RI strategy to evaluate part of MO-ERIs on-the-fly, our implementation
avoids the heavy network communication of $b_{cd}^{ab}$ 
and $\tau_{\alpha\beta}^{\gamma\eta}$ (symmetrical doubles amplitudes in AO basis 
requested by AO integral-direct algorithm~\cite{rendell:1992A,anisimov:2014A}).
With this domain-based concept, the total communication load of our approach
for the pp-Ladder term is $4N_{vir}^2N_{d}^{-1}(N_{occ}^2+2N_{ABS}+N_{occ})$ bytes. 
Although retaining the intrinsic $O(N^4)$ dependence on the problem size, the leading
cost in communication reduces significantly with respect to the number of domains $N_{d}$.

Another advantage of partially turning off symmetry is the possibility of
distributing and retrieving $b_{cd}^{\bar{a}b}$ as regular-shape blocks,
such that the tensor contraction of $L_{ij}^{\bar{a}b}$ can be realized block by block. 
At the cost of moderately increased temporary buffer per processor, 
about $4N_{vir}^2N_{occ}^2N_{pd}^{-1}$ bytes plus some arrays on the order 
of $N_{vir}^3N_{pd}^{-1}$, this permits to avoid
intensive intra-domain exchange requests of small data packages which often imposes a big challenge
for any interconnect. 



\section{Performance}
Our CCSD(T) algorithm was coded in the FHI-aims package using NAO basis sets~\cite{blum:2009A,
zhang:2013A}. In this section, we focus on the parallel performance of the CCSD code,
which is a partricularly difficult part in the massive-parallel implementation of CCSD(T), 
and has been extensively investigated. The benchmark tests were carried out on the ``HYDRA'' 
supercomputer of Max Planck Computing \& Data Facility (MPCDF). Despite HYDRA is not the most 
powerful HPC cluster available at FHI now, its hardware parameter is at the same level of 
supercomputers that were used in recent benchmark works on state-of-the-art massively parallel 
implementations of CCSD~\cite{solomonik:2014A,peng:2016A}, making it possible to compare the 
parallel performance of our code with others directly. In the appendix, we will provide the 
detailed description of HYDRA as well as the other two supercomputers, \emph{i.e.}\ BlueBidge 
and Cray XC30, that are used to produce the CCSD results for comparison in this section.

\begin{figure}
\includegraphics[width=0.5\textwidth]{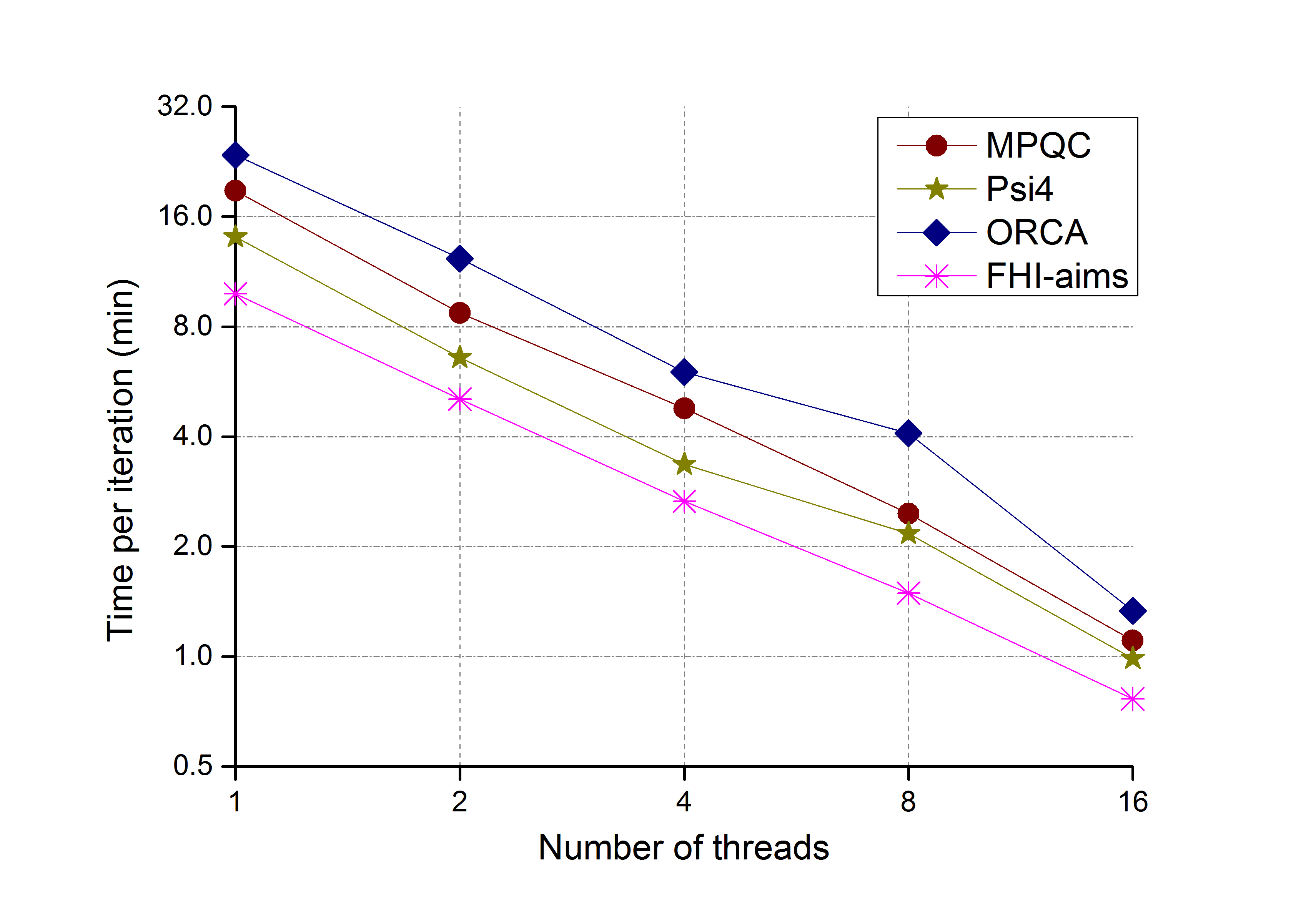}
\caption{
	\label{fig:multithreads} 
	Multi-threaded performance of the present and state-of-the-art implementations of 
	frozen-core CCSD for a (H$_2$O)$_{10}$ cluster at the cc-pVDZ level. The CCSD implementations
	on MPQC, Psi-4, and FHI-aims are using the RI approximation; while the CCSD code in ORCA uses 
	the traditional MO-only algorithm. MPQC, Psi-4, and ORCA results were taken from 
    Ref.~\cite{peng:2016A}. All these results were produced using a compute node of 2 Sandy Bridge
    CPUs with 16 physical cores in total and 64 GB of RAM.
}
\end{figure}

We first benchmarked the multi-threaded performance of the OpenMP application in our code using
a single Sandy Bridge node in HYDRA. The frozen-core RI-CCSD calculations were carried out for 
a (H$_2$O)$_{10}$ cluster with the cc-pVDZ basis set ($N_{occ}=40, N_{vir}=190, N_{ABS}=920$). 
All these calculations use the domain-based distributed-memory setting with one domain and one 
processor ($N_{d}=1,N_{pd}=1$) consistently. Figure~\ref{fig:multithreads} presents the time cost 
per CCSD iteration of present work against the number of threads (labeled as FHI-aims). Compared 
with the calculation using one thread only (9.9 minutes per CCSD iteration), our RI-based CCSD 
code provides a speedup of 12.3$\times$ on 16 threads, resulting in the cost of 0.8 minutes per 
CCSD iteration. In the recent paper of the TiledArray-based massive-parallel CCSD(T) implementation 
in MPQC~\cite{peng:2016A}, the multi-threaded performance of their code has been investigated 
together with other state-of-the-art CCSD codes on the same system (frozen-core, 
(H$_{2}$O)$_{10}$@cc-pVDZ) and the same hardware (2 Sandy Bridge CPUs with 8 physical cores per 
CPU, 64 GB of RAM per node). 
Figure~\ref{fig:multithreads} also shows the results of three codes with the best multi-threaded 
parallel performance reported in the literature, including the MO-only CCSD code in ORCA and 
two RI-based CCSD codes in MPQC and Psi-4. 
Both MO-only CCSD in ORCA and RI-CCSD in MPQC present a superlinear multi-threaded scaling, 
resulting in 18$\times$ and 16.9$\times$ speedups on 16 threads against their own performances
with one thread (it is not clear if the hyper-threading mode is active in their benchmark). 
The RI-based CCSD code in Psi-4 also possesses an excellent thread scaling with 14.5$\times$ speedup
on 16 threads.
By partially utilizing the permutation symmetry, our RI-CCSD code with one thread is about 1.9 times 
faster than MPQC in which the symmetry has been completely turned off~\cite{peng:2016A}; 
and it remains about 1.4 times faster than MPQC on threads according to the time cost 
reported in their paper.

\begin{figure}
\includegraphics[width=0.5\textwidth]{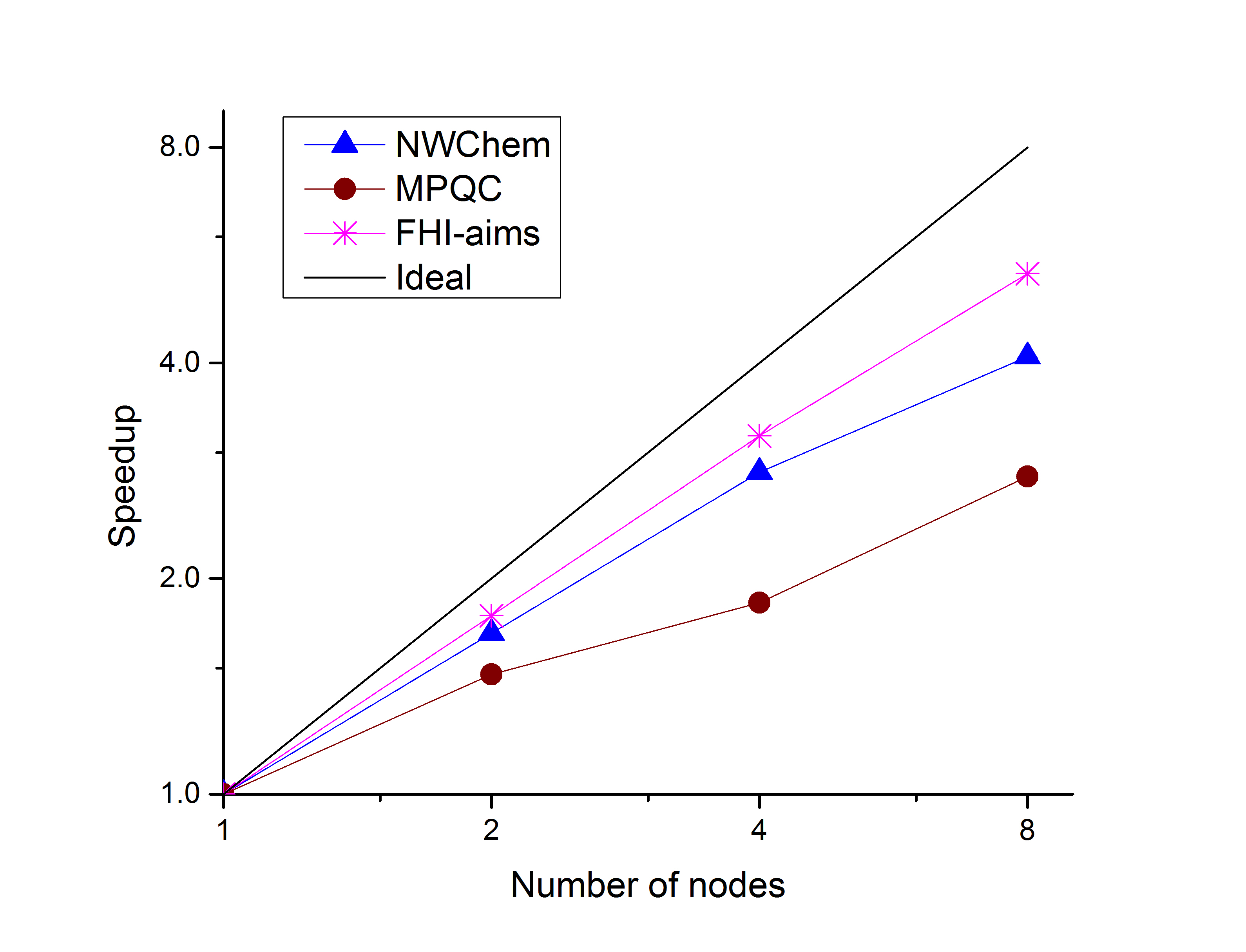}
\caption{
	\label{fig:MPIw10} 
	Parallel performance of the present and state-of-the-art CCSD codes against the number of nodes 
	for a 10-water cluster with the cc-pVDZ basis. The frozen-core CCSD calculations using FHI-aims 
	were carried out on HYDRA; while the results of MPQC and NWChem were obtained on BlueRidge using
	the same Sandy Bridge nodes as on HYDRA and reported in Ref.~\cite{peng:2016A}. Please refer
	to the appendix for a detailed description of HYDRA and BlueRidge.
}
\end{figure}

In figure~\ref{fig:MPIw10}, we then benchmarked the domain-based distributed-memory parallel 
performance of our code versus the number of Sandy Bridge nodes for the same system using the 
same basis set, \textit{i.e.}\ (H$_2$O)$_{10}$ with cc-pVDZ. As this system is small enough, 
we took each node (2 MPI processes with 8 OpenMP threads per process) as a domain ($N_{pd}=2, 
N_{d}=\textrm{the number of nodes}$). As discussed in the previous section, the domain-based 
strategy developed in this work prepares a full copy of necessary intermediate data in each 
domain, thus significantly minimizing the inter-domain communication. Therefore, it is not 
surprised to see that our code exhibits an excellent strong scaling 
against the number of nodes (or domains). The resulting parallel efficiency on 8 nodes (or 8 domains) 
is above 75\% compared to the performance on 1 node. 
On the other hand, we also investigate the parallel performance of our code using all available
nodes as a domain ($N_{d}=1,N_{pd}=N_{p}$, the standard distributed-memory strategy). 
Our algorithm still requests the intra-domain communication
in the order of $O(N^4)$, but with much smaller prefactor (see Section~\ref{section:dbdm}).
Together with the loop-blocking design in the MPI/OpenMP framework,
it enables the parallel performance of our approach with $N_{d}=1, N_{pd}=N_{p}$ to be almost identical  
to the multi-domain calculations in this system, which therefore are not plotted
in figure~\ref{fig:MPIw10}.
The same benchmark tests on two state-of-the-art CCSD implementations (MPQC TiledArray-based CCSD code 
and NWChem TCE CCSD code) were recently performed on ``BlueRidge'' cluster which are composed by 
the same Sandy Bridge nodes as HYDRA~\cite{peng:2016A}. 
A slightly different setting in their work is the use of 8 MPI processes with 2 OpenMP threads 
in each node. Taking the results reported in Peng's paper~\cite{peng:2016A}, figure~\ref{fig:MPIw10} 
also plots the strong-scaling behaviors of MPQC and NWChem in this benchmark. 
Using 1 node, the time costs per CCSD iteration are 50 seconds and 245 seconds for MPQC and 
NWChem, respectively.
As the intermediate data were distributed uniformly across all processes
in both implementations, the reported parallel efficiency is around 35\% to 50\% from 1 node to 
8 nodes.

\begin{figure}
	\includegraphics[width=0.5\textwidth]{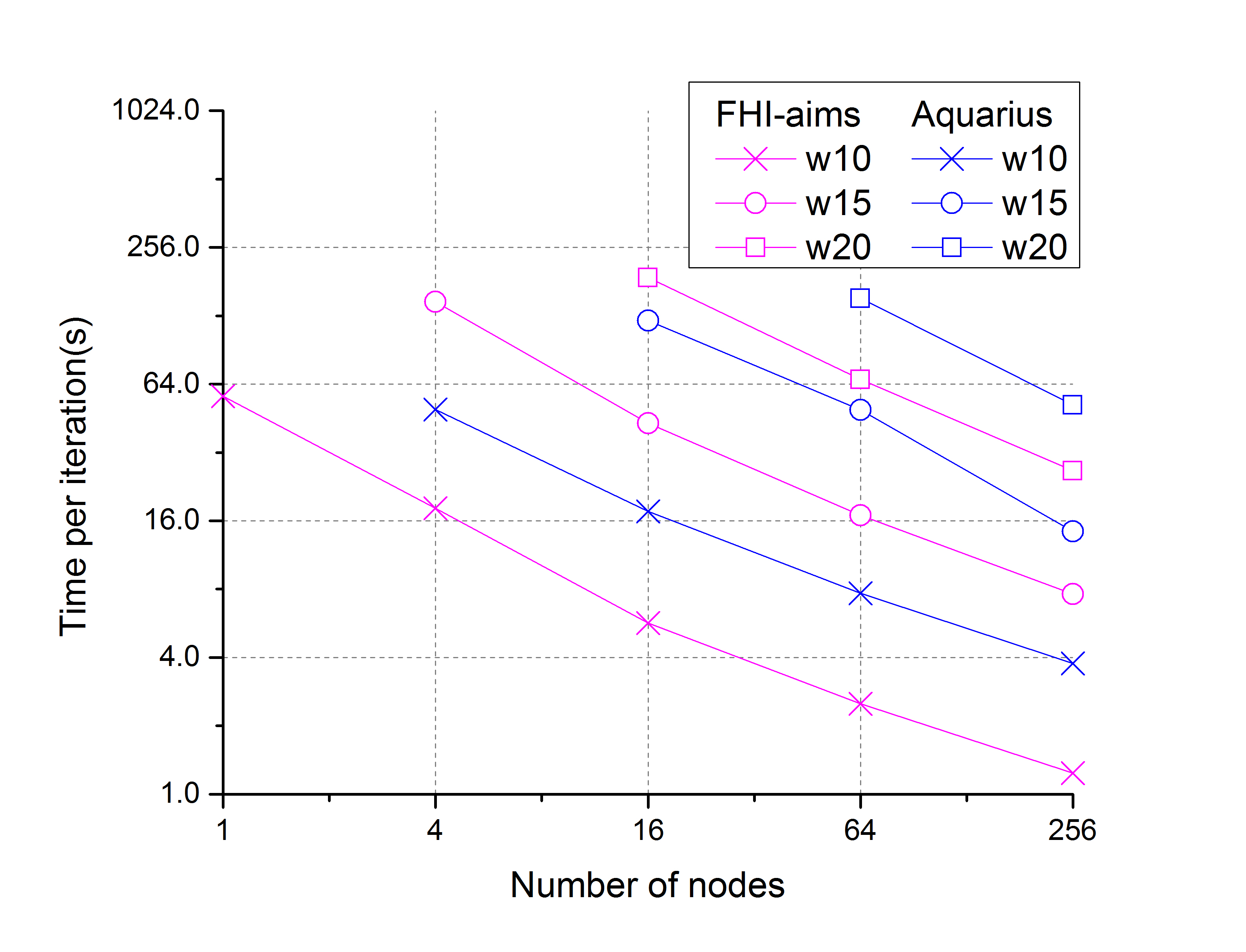}
	\caption{
		\label{fig:CTFWn} 
		All-electron CCSD strong scaling of water clusters with cc-pVDZ. The calculations of 
		FHI-aims were carried out on HYDRA using Ivy Bridge nodes, while those CCSD results of 
		CTF-based Aquarius code were obtained on Cray XC30 using Ivy Bridge nodes as well and 
		reported on Ref.~\cite{solomonik:2014A}. The comparison of HYDRA and Cray XC30 
		supercomputers can be found in the appendix.
}
\end{figure}

Figure~\ref{fig:CTFWn} presents the strong scaling of all-electron CCSD calculations for water 
clusters with different sizes ((H$_2$O)$_{n}$ with $n=10,15,\textrm{ and }20$) on HYDRA up to 256 
Ivy Bridge nodes equipping 64 GB of RAM per node and 5120 physical cores in total. 
The cc-pVDZ basis set was employed. 
In this benchmark, we set the 
number of processors per domain $N_{pd}=2,8,\textrm{ and }32$ for the clusters with 10, 15, and 20 
waters, respectively. 
In line with the above observation, figure~\ref{fig:CTFWn} demonstrates again that our domain-based 
distributed-memory strategy enables CCSD calculations to scale on many thousands of cores, 
while achieving a high degree of efficiency in computation, communication, and storage.
For comparison, figure~\ref{fig:CTFWn} also plots the parallel performance of the CCSD code in
Aquarius based on the results reported in Ref.~\cite{solomonik:2014A}. Due to the use of the CTF library
to automatically manage tensor blocking, redistribution, and contractions in CC theory, their CTF-based
CCSD implementation in the standard distributed-memory framework achieves high parallel scalability on 
Cray XC30 supercomputer which is composed of Intel Ivy Bridge Xeon E5-2595 CPUs with 24 physical 
cores and 64 GB per node. Because the Aquarius CTF-based CCSD code stores all intermediate data, 
including the most consuming $v_{cd}^{ab}$, in memory, their benchmark tests were performed starting 
from 4 nodes, 16 nodes, and 64 nodes for (H$_2$O)$_{10}$, (H$_2$O)$_{15}$, and (H$_2$O)$_{20}$, 
respectively. Note that the Ivy Bridge nodes used in Cray XC30 are more powerful than those in 
HYDRA (refer to the appendix for more details).
However, the time per CCSD iteration for Aquarius is significantly longer, which is most likely due 
to the use of the open-shell CCSD formulism in the Aquarius code; and thus more FLOPs are needed.

\begin{figure}
	\begin{tikzpicture}[remember picture]
		\node[text width=8.8cm, align=center] (d00) {
		\includegraphics[width=\textwidth]{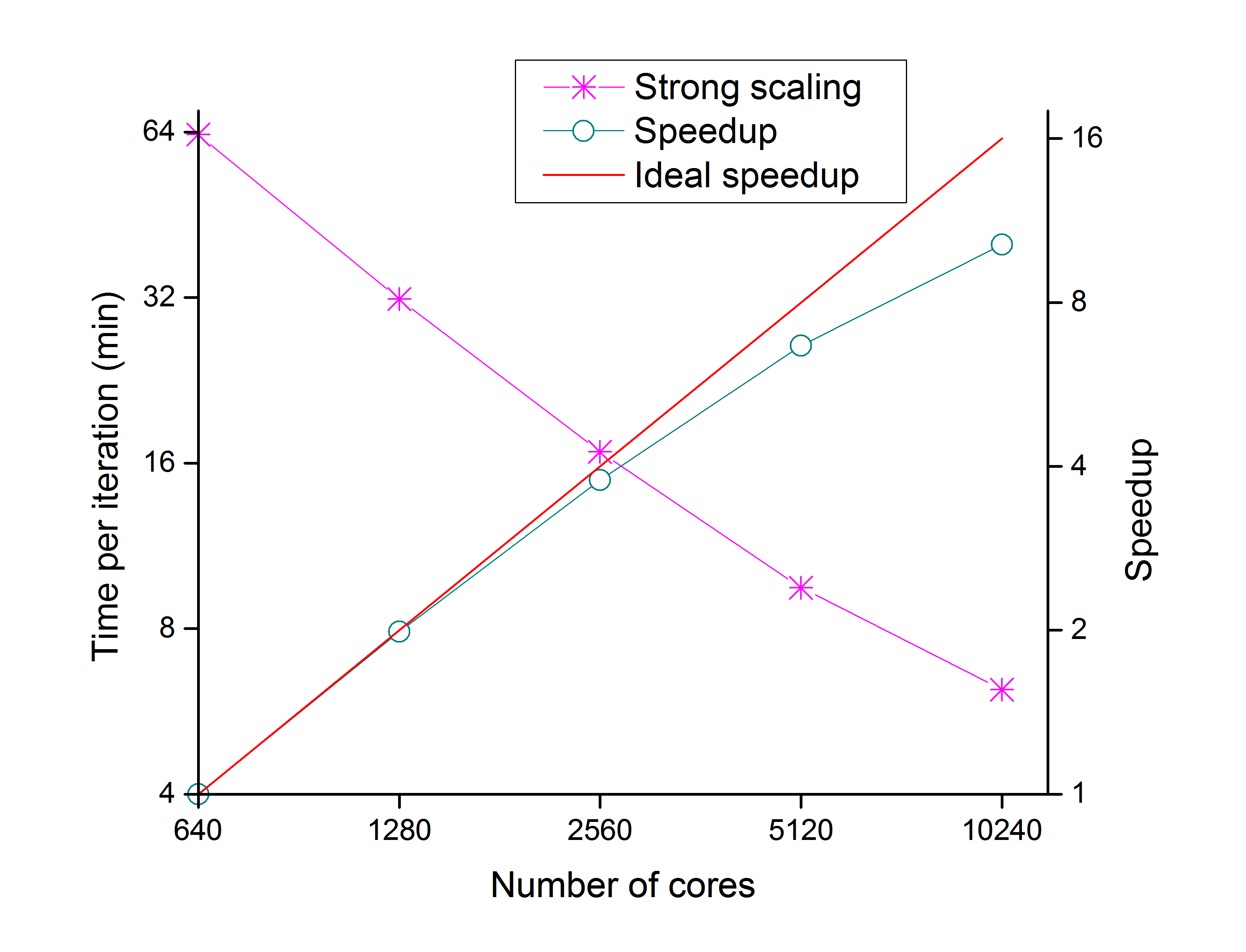}};
		\node[text width=6.0cm, rotate=40, opacity=0.3, align=center, below left=-6.3cm and -6.3cm of d00] {
		\includegraphics[width=\textwidth]{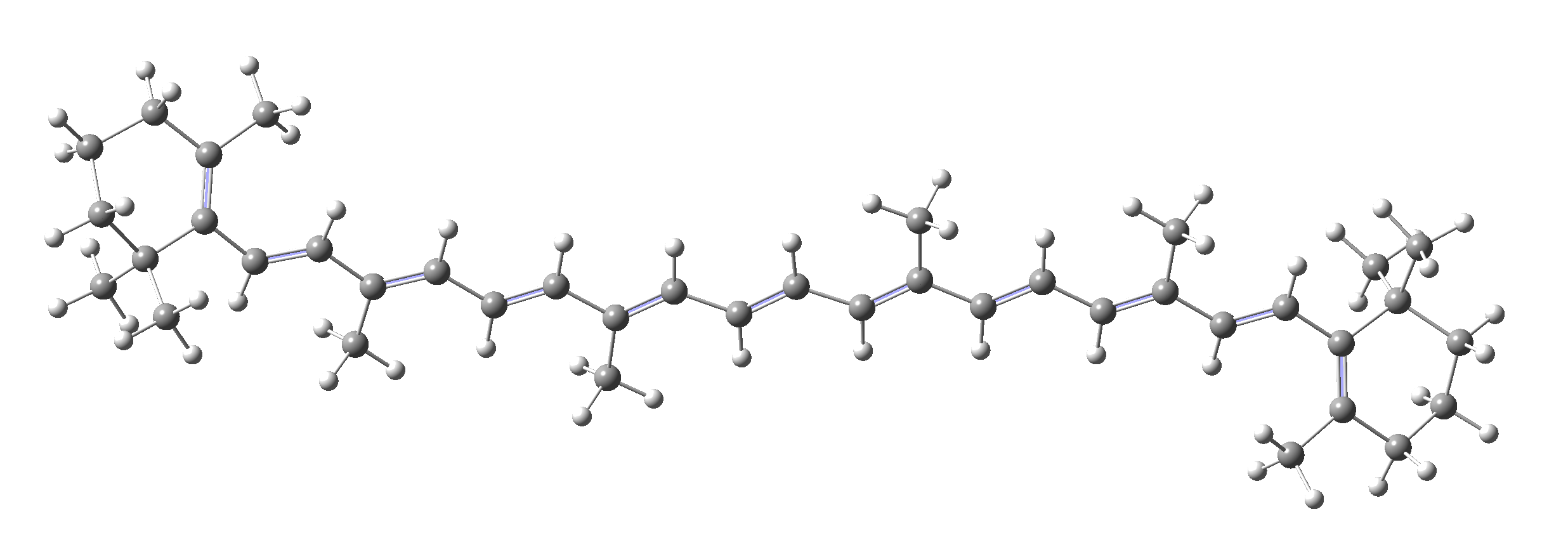}};
	\end{tikzpicture}
\caption{\label{fig:carotene} CCSD strong scaling of $\beta$-carotene with mTZ basis set on HYDRA.}
\end{figure}

As one of the largest systems used in the benchmark of state-of-the-art CCSD 
implementations~\cite{peng:2016A,hu:2014A}, the $\beta$-carotene molecule with 96 atoms was 
also investigated in this work. 
The frozen-core CCSD calculation 
with a modified TZ basis set involves 108 valence orbitals, 884 unoccupied orbitals, and 
4504 auxiliary basis functions. 
According to the memory requirement to store a full copy of intermediate data in our 
domain-based distributed-memory strategy, 
32 Ivy Bridge fat nodes (128 GB of RAM per node; 640 physical cores in total) are grouped as a domain
with the number of processors per domain $N_{pd}=64$.
Figure~\ref{fig:carotene} presents the strong-scaling performance of our code in this case.
Using 640 cores as one domain, the time per CCSD iteration costs 63 minutes, which can be reduced
to only 6 minutes if 10240 cores are employed ($N_{pd}=64$, $N_{d}=16$), resulting in the parallel efficiency
as high as 66\%. For comparison, the time per CCSD iteration reported for the AO integral-direct CCSD code in MPQC
is about 100 minutes on 32 Sandy Bridge nodes with 512 cores~\cite{peng:2016A}; while it costs about 115 minutes 
for the NWChem TCE-based CCSD code on 48 Sandy Bridge nodes with 768 cores~\cite{hu:2014A}.

Table~\ref{table:T} summarizes the time consumption of CCSD and (T) calculations for water clusters
by our domain-based distributed-memory implementation in FHI-aims.
As the computational cost of (T) in term of FLOP is one order of magnitude higher than the CCSD procedure,
it often spends longer time on the (T) part.
As shown in table~\ref{table:T}, the cost ratio of (T) to CCSD is about 5 to 6 for the 10-water cluster, 
which, however, increases quickly to above 40 for 20 waters. 
Luckly, the massive-parallel implementation of the (T) part can be very efficient, as demonstrated in
table~\ref{table:T}. 

\begin{table}[h]
	\begin{ruledtabular}
		\caption{Time consumption (unit minute) of frozen-core CCSD and (T) calculations for (H$_2$O)$_{n}$ with 
	$n=10,15,\textrm{ and } 20$ with cc-pVDZ basis set. 
	Each Sandy Bridge (or Ivy Bridge) node executes 2 MPI processors with 8 (or 10) OpenMP threads.}
\label{table:T}
\begin{tabular}{ccccc}
	Complex     & $N_{pd}$
	\footnote{Calculations of (H$_2$O)$_{10}$ are all on Sandy Bridge nodes; while those of 
		(H$_2$O)$_{15}$ and (H$_2$O)$_{20}$ are on
	Ivy Bridge nodes.}
	& $N_{d}$ & CCSD
	\footnote{For simplicity, we assume that CCSD calculations need 12 iterations to converge for all three molecules.} & (T) \\
\hline
\multirow{4}{*}{(H$_2$O)$_{10}$} &  2       &  1      &  8.7 &  49.6   \\
                                 &  2       &  2      &  4.6 &  25.1   \\
                                 &  2       &  4      &  2.6 &  12.6   \\
				&  2       &  8      &  1.4 &   6.5   \\
								 \hline
\multirow{3}{*}{(H$_2$O)$_{15}$} &  4       &  2      & 16.4 & 220.1   \\
                                 &  4       &  4      &  9.1 & 109.1   \\
                                 &  4       &  8      &  5.5 &  54.9   \\
								 \hline
\multirow{3}{*}{(H$_2$O)$_{20}$} & 16       &  2      & 23.5 & 691.6  \\
                                 & 16       &  4      & 12.9 & 358.5   \\
                                 & 16       &  8      &  7.5 & 183.3   \\
\end{tabular}
\end{ruledtabular}
\end{table}

\section{Results and Discussions}
\begin{figure}
\includegraphics[width=0.5\textwidth]{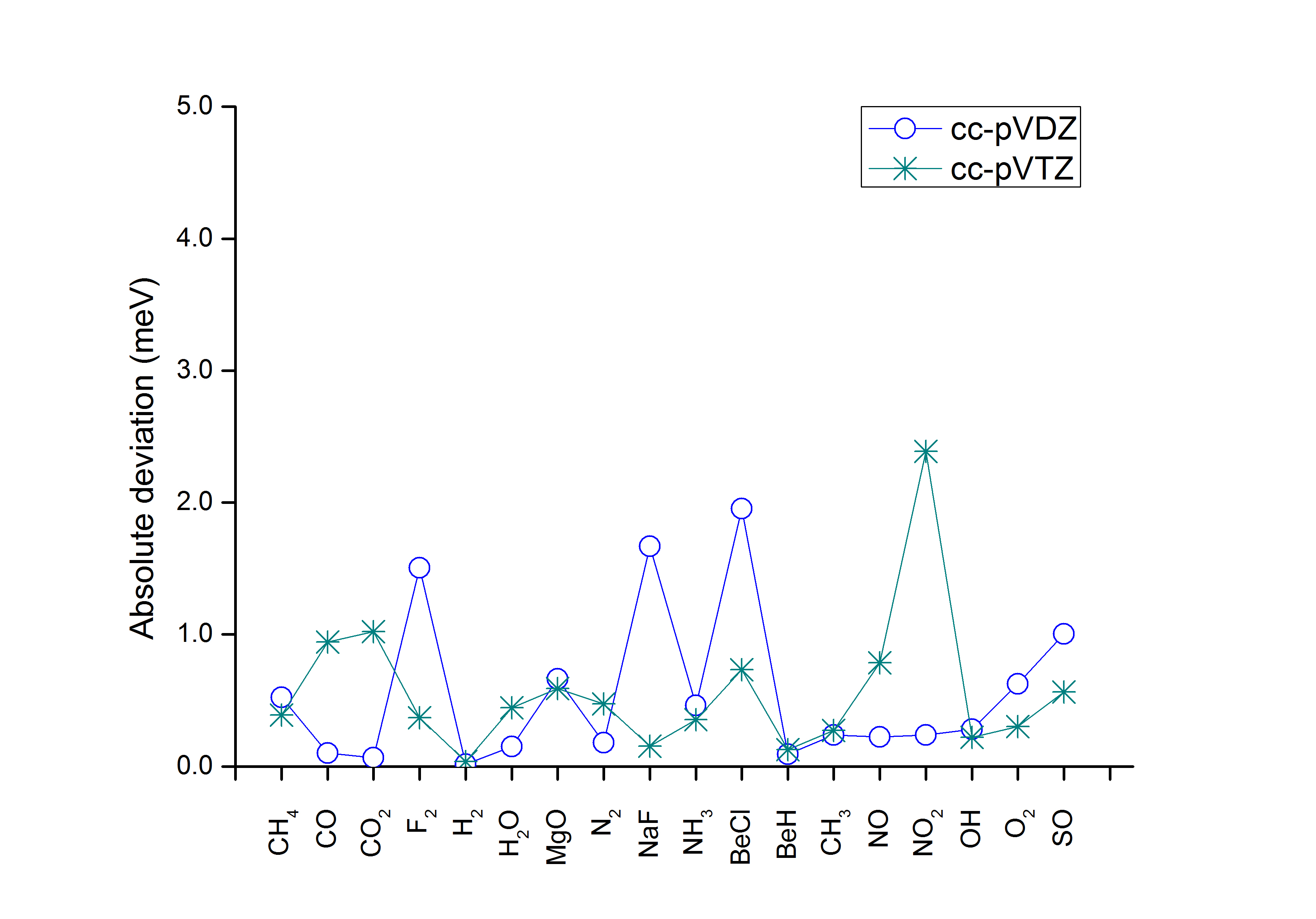}
\caption{\label{fig:AE} Absolute deviation of frozen-core RI-CCSD total energies for a test set of 18 small molecules.
from analytic integral implementation. Two sets of references data with cc-pVDZ and cc-pVTZ basis sets were calculated 
by the GAMESS package with analytic MO-ERIs~\cite{schmidt:1993A,olson:2007A}.}
\end{figure}

As the RI approximation is used to evaluate MO-ERIs~\cite{ren:2012B,ihrig:2015A}, it is
necessary to examine the numerical accuracy of our RI-CC approaches compared with the CC results
using analytic MO-ERIs and the same basis set.
Taking the frozen-core CCSD total energies obtained from GAMESS with analytic MO-ERIs as reference,
Figure~\ref{fig:AE} shows the absolute deviation of our RI-CCSD results against a test set of 18 
small molecules and radicals, including \texttt{CH$_{4}$, CO, CO$_{2}$, F$_{2}$, H$_{2}$, H$_{2}$O, MgO, 
N$_{2}$, NaF, NH$_{3}$, BeCl, BeH, CH$_{3}$, NO, NO$_{2}$, OH, O$_{2}$, \textrm{and} SO}. 
For both Gaussian-type basis sets, cc-pVDZ and cc-pVTZ, the absolute deviations of our RI-CCSD results can
be smaller than 3 meV. The mean absolute deviations (MADs) are less than 0.6 meV for both basis sets, 
demonstrating that the auxiliary basis set used in FHI-aims is accurate enough for the calculations of CC methods.

The CCSD(T) method has been widely used to produce accurate theoretical reference data of popular 
molecular test sets, for example the S22 set with 22 bio-oriented non-covelent 
interactions~\cite{Jurecka:2006} and CYCONF with 10 relative energies of cysteine 
conformations~\cite{Wilke:2009}. These test sets have been widely used to benchmark newly developed 
electronic-structure approaches, mainly in density functional theory.
However, due to the demanding computational cost and the slow basis-set convergence, it is
very difficult (often unfeasible under today's computer capacity) 
to provide CCSD(T) results in complete basis-set (CBS) limit for large molecules in these test sets. 
In this work, we utilize a combination methodology to approach the CBS CCSD(T) results
\begin{equation}
	\label{eq:cbs}
	E_{CBS}^{\textrm{CCSD(T)}}\approx E_{CBS}^{\textrm{MP2}}+\Delta E_{finite}^{\textrm{CCSD(T)}},
\end{equation}
where the couple-cluster correction at a finite basis set $\Delta E_{finite}^{\textrm{CCSD(T)}}$ 
is defined as,
\begin{equation}
	\Delta E_{finite}^{\textrm{CCSD(T)}}=E_{finite}^{\textrm{CCSD(T)}}-E_{finite}^{\textrm{MP2}},
\end{equation}
and the converged MP2 total energy $E_{CBS}^{\textrm{MP2}}$
\begin{equation}
	E_{CBS}^{\textrm{MP2}}=E_{CBS}^{\textrm{HF}}+E_{CBS}^{\textrm{MP2,c}}
\end{equation}
is achieved by two-point extrapolations for the converged HF total energy $E_{CBS}^{\textrm{HF}}$
and the MP2 correlation energy $E_{CBS}^{\textrm{MP2,c}}$ using the correlation-consistent basis set,
\begin{equation}
	E_{nZ}^{\textrm{HF}}\left(n\right)=E_{CBS}^{\textrm{HF}}+Ae^{-\alpha n},
\end{equation}
\begin{equation}
	E_{nZ}^{\textrm{MP2,c}}\left(n\right)=E_{CBS}^{\textrm{MP2,c}}+Bn^{-3}.
\end{equation}
Here, $n$ denotes the cardinal number of correlation-consistent basis sets, for example the popular 
cc-pV$n$Z ($n=D,T,Q\textrm{ and } 5$) and NAO-VCC-$nZ$ with $n=2,3,4,\textrm{ and }5$ provided in 
FHI-aims~\cite{zhang:2013A}. This combination strategy was proposed based on the assumption that
the energy difference between MP2 and CCSD(T) $\Delta E_{finite}^{\textrm{CCSD(T)}}$ converges 
much faster against the basis-set size than the MP2 and CCSD(T) energies themselves; and
more accurate $E_{CBS}^{\textrm{CCSD(T)}}$ can be obtained if the finite basis set used
for $\Delta E_{finite}^{\textrm{CCSD(T)}}$ is larger.
In this work, we chose $\alpha=1.63$ for $E_{CBS}^{\textrm{HF}}$ as recommended by Halkier 
\textit{et.\ al.}~\cite{Halkier:1999}. 
$E_{CBS}^{\textrm{MP2,c}}$ was obtained by the two-point extrapolation between NAO-VCC-4Z and 5Z; 
while the calculations of $\Delta E_{finite}^{\textrm{CCSD(T)}}$ were carried out with NAO-VCC-3Z or 4Z according
to the molecular size in calculations.

The S22 test set was constructed by Jura\v{c}ka \textit{et.\ al.}\ in 2006~\cite{Jurecka:2006} to benchmark
the accuracy of theoretical methods in the description of non-covalent interactions.
The CCSD(T) reference data reported in the original paper, designated S22-RefA, were calculated following 
a similar combination strategy (equation~\ref{eq:cbs}) with the CCSD(T) correction 
$\Delta E_{finite}^{\textrm{CCSD(T)}}$ evaluated
with smaller Gaussian-type basis sets (6-31G** and cc-pVDZ). In the past years, there were a number of literatures
to update the CCSD(T) reference data for the S22 test set using larger basis sets for both extrapolation and correction
terms in equation~\ref{eq:cbs}~\cite{Takatani:2010,Rodeszwa:2010,Marshall:2011}. The revised CCSD(T) reference data
provided by Sherrill's group in 2012 are considered to be the most accurate results to date, designated S22-RefB.
However, these CCSD(T) reference data reported are all based on Gaussian-type basis sets, mainly (aug)-cc-pV$n$Z.

Table~\ref{table:S22} lists 22 non-interaction energies of CCSD(T) quality calculated by our RI-CCSD(T) code with 
NAO-VCC-$n$Z. Compared with the up-to-date reference data S22-RefB, the present work provides the deviations usually 
smaller than 0.1 kcal/mol, with the MAD of 0.059 kcal/mol and the maximum deviation of 0.19 kcal/mol in the 
parallel displaced benzene dimer (No.\ 11); while the MAD between S22-RefB and S22-RefA is 0.136 kcal/mol.
Our data demonstrates that accurate CCSD(T) results can be obtained by using our RI-CCSD(T) code with the 
correlation-consistent NAO basis sets, NAO-VCC-$n$Z. 

\begin{table}[h]
\begin{ruledtabular}
	\caption{CCSD(T)/CBS weak interaction energies for the S22 test set (kcal/mol). The 22 test cases are organized
	in the same order as original~\cite{Jurecka:2006}.}
\label{table:S22}
\begin{tabular}{rrrrr}
Index &  Present &   S22-RefA      &    S22-RefB    \\
\hline
1     &  -3.13   &   -3.17         &      -3.13     \\
2     &  -4.98   &   -5.02         &      -4.99     \\
3     &  -18.78  &   -18.61        &      -18.75    \\
4     &  -16.14  &   -15.96        &      -16.06    \\
5     &  -20.73  &   -20.65        &      -20.64    \\
6     &  -16.99  &   -16.71        &      -16.93    \\
7     &  -16.70  &   -16.37        &      -16.66    \\
8     &  -0.52   &    -0.53        &      -0.53     \\
9     &  -1.53   &    -1.51        &      -1.47     \\
10    &  -1.47   &    -1.50        &      -1.45     \\
11    &  -2.84   &    -2.73        &      -2.65     \\
12    &  -4.38   &    -4.42        &      -4.26     \\
13    &  -9.74   &    -10.12       &      -9.81     \\
14    &  -4.67   &    -5.22        &      -4.52     \\
15    &  -11.82  &    -12.23       &      -11.73    \\
16    &   -1.54  &    -1.53        &      -1.50     \\
17    &   -3.29  &    -3.28        &      -3.28     \\
18    &   -2.35  &    -2.35        &      -2.31     \\
19    &   -4.55  &    -4.46        &      -4.54     \\
20    &   -2.79  &    -2.74        &      -2.72     \\
21    &   -5.65  &    -5.73        &      -5.63     \\
22    &   -7.06  &    -7.05        &      -7.10     \\
\end{tabular}
\end{ruledtabular}
\end{table}
 
We then applied our code to calculate the CCSD(T) reference data of 10 relative energies of cysteine 
conformers in the CYCONF test set~\cite{Wilke:2009}. The geometries of all 11 stationary conformers 
were optimized at MP2/cc-pVTZ level. The CCSD(T)/CBS reference data provided in the original literature, 
denoted as CYCONF-RefA,
were obtained following the similar combination strategy (equation~\ref{eq:cbs}) with the CCSD(T) correction 
$\Delta E_{finite}^{\textrm{CCSD(T)}}$ at the basis-set level of cc-pVTZ. In the present work, we evaluated 
the CCSD(T) correction using a NAO basis set with more basis functions, NAO-VCC-4Z. 
Table~\ref{table:CYCONF} summarizes the relative energies with respect to the most stable conformer (No.\ 1).
It can be seen that our results predict the order of conformers in terms with the relative energies, which is the same
as those by CYCONF-RefA~\cite{Wilke:2009}. However, inspecting table~\ref{table:CYCONF} reveals that the 
CCSD(T)/CBS relative energies of the present work are systematically smaller. To further study this 
systematic deviation, we suggested to update the original reference data with a better CCSD(T) correction 
$\Delta E_{finite}^{\textrm{CCSD(T)}}$ at the basis-set level of (aug)-cc-pVQZ.

\begin{table}[h]
	\caption{CCSD(T)/CBS relative energies of gaseous cysteine obtained in the present work and from 
	the original paper, CYCONF-RefA~\cite{Wilke:2009} (kcal/mol).}
\label{table:CYCONF}
\begin{ruledtabular}
\begin{tabular}{rrr}
       Index         &     Present        &   CYCONF-RefA\\
\hline
          1          &      0.00          &        0.00  \\
          2          &      1.46          &        1.52  \\
          3          &      1.54          &        1.61  \\
          4          &      1.87          &        1.95  \\
          5          &      1.73          &        1.80  \\
          6          &      2.01          &        2.10  \\
          7          &      1.85          &        1.93  \\
          8          &      2.12          &        2.18  \\
          9          &      2.31          &        2.36  \\
         10          &      2.60          &        2.56  \\
         11          &      2.61          &        2.67  \\
\end{tabular}
\end{ruledtabular}
\end{table}

\begin{table}[h]
	\begin{ruledtabular}
		\caption{CCSD(T)/CBS isomerization energies obtained in the present work together 
		with the experimental data for 34 organic reactions in the ISO34 set~\cite{Grimme:2007}.}
\label{table:ISO34}
\begin{tabular}{rrrrrr}
	Index    &   Present &   Expt.   &   Index    &    Present    &  Expt.  \\
\hline
     1         &   1.58  &    1.62  &      18       &  11.69   &    11.16  \\
     2         &   23.30 &   21.88  &      19       &   4.59   &    0.00   \\
     3         &   7.44  &    7.20  &      20       &  17.95   &    20.23  \\
     4         &   1.07  &    0.99  &      21       &   1.15   &    0.94   \\
     5         &   1.22  &    0.93  &      22       &   3.27   &    3.23   \\
     6         &   2.39  &    2.62  &      23       &   5.49   &    5.26   \\
     7         &  11.10  &    11.15 &      24       &  11.98   &   12.52   \\
     8         &  22.71  &    22.90 &      25       &  25.98   &   26.49   \\
     9         &  6.42   &    6.94  &      26       &  16.68   &   18.16   \\
    10         &  3.73   &    3.58  &      27       &  65.30   &   64.17   \\
    11         &  1.50   &    1.91  &      28       &  30.89   &   31.22   \\
    12         &  44.90  &    46.95 &      29       &  13.05   &   11.90   \\
    13         &  36.46  &    36.04 &      30       &   9.89   &   13.60   \\
    14         &  24.18  &    21.30 &      31       &  15.23   &   14.05   \\
    15         &  8.21   &    7.26  &      32       &   6.56   &    2.40   \\
    16         &  10.12  &    10.81 &      33       &   8.44   &    5.62    \\
    17         &  28.40  &    26.98 &      34       &   6.59   &    7.26    \\
\end{tabular}
\end{ruledtabular}
\end{table}

\begin{figure}
\includegraphics[width=0.5\textwidth]{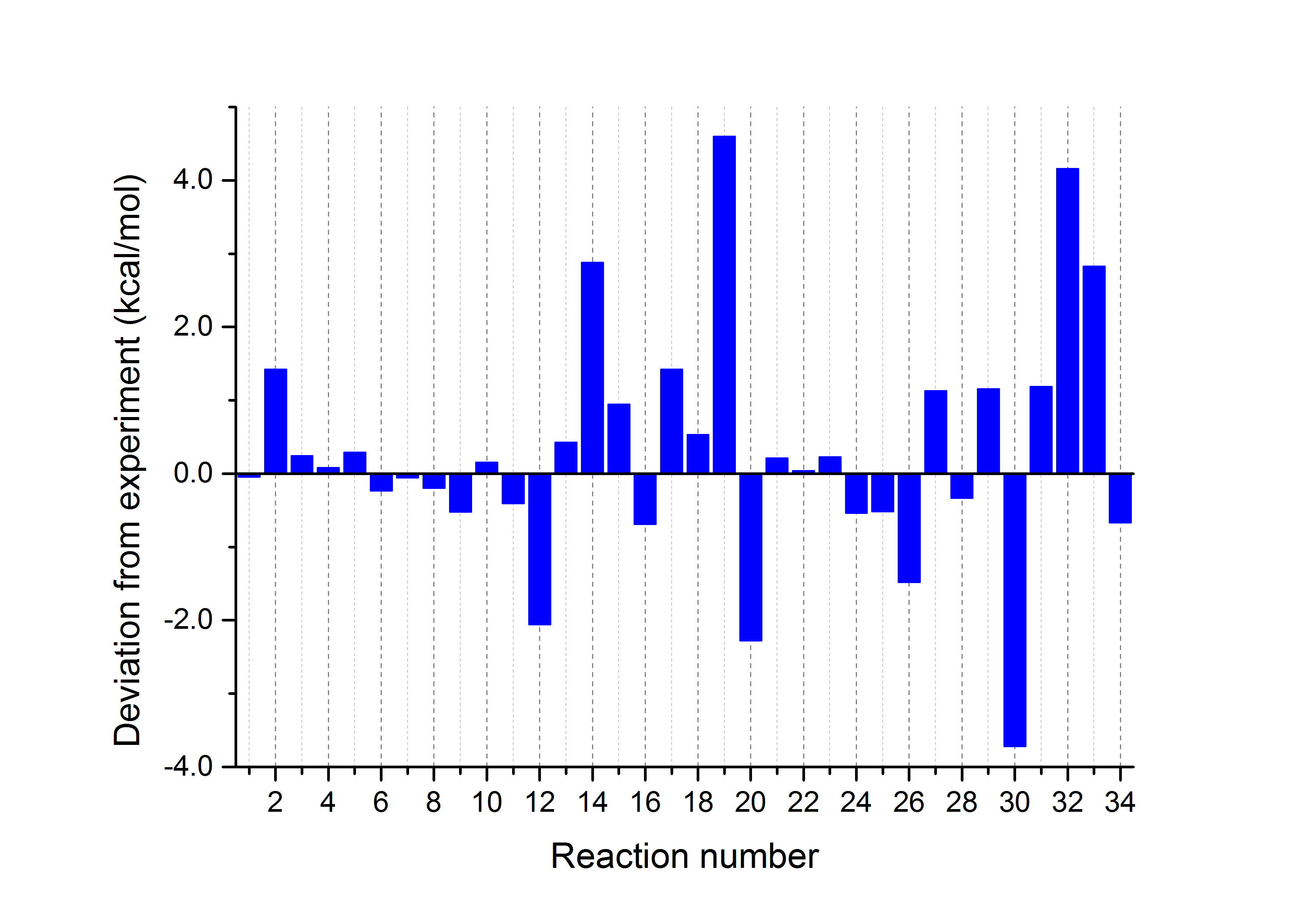}
\caption{\label{fig:ISO34} Deviations of CCSD(T)/CBS results against the experimental data for ISO34 (kcal/mol).}
\end{figure}

Isomerization is a well-defined reaction process in organic chemistry. 
The ISO34 test set is composed of 34 isomerization energies of small organic molecules~\cite{Grimme:2007}.
The experimental reference data provided in the seminal paper are presented in table~\ref{table:ISO34}.
In this work, we produced an accurate theoretical reference data at the CCSD(T)/CBS level, in which
the CCSD(T) correction $\Delta E_{finite}^{\textrm{CCSD(T)}}$ was evaluated with the NAO-VCC-3Z basis set.
We also presented the CCSD(T)/CBS reference data in table~\ref{table:ISO34} and visualized the deviations
of CCSD(T)/CBS data against experimental data in figure~\ref{fig:ISO34}. Despite the MAD between theoretical
and experimental data is 1.1 kcal/mol, approaching the expected 'chemical accuracy', it can be seen that there are
13 reactions of which the deviation is larger than 1.0 kcal/mol with the maximum deviation of 4.59 kcal/mol
occurring at the 19th reaction. For the sake of benchmarking newly developed electronic-structure methods, 
we suggest to use the CCSD(T)/CBS reference data, so that the comparison can be based on exactly the 
same molecular geometry and immune to the experimental uncertainty, making the comparison well-defined.

\section{Conclusions}
In this work, we introduce a domain-based distributed-memory strategy to implement a massive-parallel
CCSD(T) code in the NAO framework for molecules. In this model, the compute processors are grouped into
a number of domains. As each domain possesses a full copy of all intermediate data, 
the CCSD(T) calculations can be carried out in each domain with slight inter-domain communications.
The permutation symmetry is partially turned off in our algorithm, and the RI approximation
are used to evaluate $v_{cd}^{ab}$ and part of $v_{cd}^{ia}$ on-the-fly. These choices result in 
an efficient parallel algorithm with optimal intra-domain communication. We demonstrate that our RI-CCSD(T)
implementation in FHI-aims exhibits an outstanding parallel performance, which is scalable from a multi-threaded
calculations in one compute node to 512 nodes with 10240 CPU cores.

As the first implementation of CCSD(T) in the NAO framework, we demonstrate that the numerical
error due to the use of RI approximation in our RI-CCSD(T) code can be negligible. Together with
the correlation-consistent NAO basis sets, NAO-VCC-$n$Z, we produce the CCSD(T)/CBS reference data
for three popular test sets in quantum chemistry, including S22, CYCONF, and ISO34. 
Our CCSD(T)/CBS results 
are in good agreement with the theoretical reference data obtained using Gaussian-type basis sets for
S22 and CYCONF. 
To replace the experimental reference data for ISO34, we suggest the use of our CCSD(T)/CBS results
in the future methodology development and benchmark.
\section{Acknowledgments}
IYZ is grateful to the support from the 14th Recruitment Program of Young Professionals in China.

\section{Appendix}
The ``HYDRA'' supercomputer of MPCDF was used to produce the CCSD(T) results of the FHI-aims 
code in this work. 610 nodes of HYDRA have 2 Intel Sandy Bridge Xeon 
E5-2670 central processing units (CPUs) with 8 physical cores per CPU and 
3500 nodes have 2 Ivy Bridge E5-2680 CPUs with 10 physical cores per CPU. 
Most of the compute nodes in HYDRA have 64 GB RAM; while 200 of the Ivy Bridge nodes and 20 of 
the Sandy Bridge nodes equip a large RAM of 128 GB, namely fat nodes. 
The cross-node connection of HYDRA goes by a fast InfiniBand FDR14 network.
FHI-aims was compiled with Intel Parallel Studio XE 16.0 and Intel MPI 5.1 
along with parallel MKL 11.3 and Intel OpenMP.

The ``BlueRidge'' supercomputer of the Virginia Tech research communicty contains 
408 nodes with 2 Sandy Bridge Xeon E5-2670 CPUs and 64 GB (or 128 GB) RAM 
per node~\cite{peng:2016A}, which is thus the same as the Sandy Bridge nodes of 
HYDRA introduced above. BlueRidge is connected via InfiniBand as well.
The CCSD results of MPQC, Psi4, and ORCA in figure~\ref{fig:multithreads} and those of 
NWChem and MPQC in figure~\ref{fig:MPIw10} were produced by using the Sandy Bridge nodes 
on BlueRidge. For comparison, the FHI-aims results shown in these figures are produced 
by using the Sandy Bridge nodes of HYDRA.

The Edison supercomputer ``Cray XC30'' contains 5586 nodes, in which there are two 12-core
Ivy Bridge E5-2695v2 CPUs and 64 GB per node. Cray XC30 equips Cray Aries high-speed
interconnect with Dragonfly topology~\cite{solomonik:2014A}. The CCSD results of CTF-based 
Aquarius code shown in figure~\ref{fig:CTFWn} were obtained by using Ivy Bridge nodes on 
Cray XC30. For comparison, the FHI-aims results in the same figure were calculated on HYDRA 
using less powerful Ivy Bridge nodes with two 10-core E5-2680 CPUs per node.

\bibliography{paper}
\end{document}